\documentclass[
a4paper,
aps,
prd,
twocolumn,
tightenlines,
preprintnumbers,
nofootinbib,
showkeys,
superscriptaddress
]{revtex4-1}

\pdfoutput=1 
\usepackage{XCharter}
\usepackage[T1]{fontenc}

\usepackage{fullpage}
\usepackage{amsfonts}
\usepackage{amsmath}
\usepackage{slashed}
\usepackage{amssymb}
\usepackage{graphicx}
\usepackage{epic}
\usepackage{eepic}
\usepackage{epsfig}
\usepackage{latexsym}
\usepackage[dvipsnames]{xcolor}
\usepackage{float}
\usepackage{multirow}
\usepackage[breaklinks]{hyperref}
\usepackage{enumitem}
\hypersetup{colorlinks=true,citecolor=red,linkcolor=NavyBlue,urlcolor=NavyBlue}
\usepackage[caption=false]{subfig}
\usepackage{natbib}
\usepackage{relsize}
\usepackage[left=2.cm,right=2.cm,top=2.25cm,bottom=2.75cm]{geometry}
\usepackage{makecell}

\newcommand{\ubr}[1]{\raisebox{1.5ex}{\hspace{#1ex}$\frown$\relax}}

\usepackage{bm}
\usepackage{subfig}

\usepackage{shorthand}
\usepackage{mciteplus}

\begin{document}
\relscale{1.05}

\title{Addressing the $ \mathbf{\textit{R}}_{\mathbf{\textit{D}}^{(*)}}$ anomalies with an $\mathbf{\textit{S}_1}$ leptoquark from $\mathbf{SO(10)}$ grand unification}

\author{Ufuk Aydemir}
\email{uaydemir@ihep.ac.cn}
\affiliation{Institute of High Energy Physics, Chinese Academy of Sciences, Beijing 100049, P. R. China}
\affiliation{School of Physics, Huazhong University of Science and Technology, Wuhan, Hubei 430074, P. R. China}

\author{Tanumoy Mandal}
\email{tanumoy.mandal@physics.uu.se}
\affiliation{Department of Physics and Astronomy, Uppsala University, Box 516, SE-751 20 Uppsala, Sweden}
\affiliation{Department of Physics and Astrophysics, University of Delhi, Delhi 110 007, India}

\author{Subhadip Mitra}
\email{subhadip.mitra@iiit.ac.in}
\affiliation{Center for Computational Natural Sciences and Bioinformatics, International Institute of Information Technology, Hyderabad 500 032, India}

\date{\today}

\begin{abstract}\noindent
Motivated by the $R_{D^{(*)}}$ anomalies, we investigate an $\mathrm{SO}(10)$ grand unification scenario 
where a charge $-1/3$ scalar leptoquark ($S_1$) remains as the only new physics candidate at the TeV scale. 
This leptoquark along with the Standard Model (SM) Higgs doublet originates from the same ten-dimensional 
real scalar multiplet in the $\mathrm{SO}(10)$ grand unification framework taking its mass close to the 
electroweak scale. We explicitly show how the gauge coupling unification is achieved with only one 
intermediate symmetry-breaking scale at which the Pati-Salam gauge group is broken into the SM group. 
We investigate the phenomenological implications of our scenario and show that an $S_1$ with a specific 
Yukawa texture can explain the $R_{D^{(*)}}$ anomalies. We perform a multiparameter scan considering 
the relevant flavour constraints  on $R_{D^{(*)}}$,  $F_L{(D^*)}$, $P_\tau(D^*)$ and $R_{K^{(*)}}^{\nu\nu}$ 
as well as the constraint coming from the $Z\to\tau\tau$ decay and the latest $\tau\tau$ resonance search 
data at the LHC. Our analysis shows that a single leptoquark solution to the observed 
$R_{D^{(*)}}$ anomalies with $S_1$ is still a viable solution.
\end{abstract}

\keywords{$B$-decay anomalies, scalar leptoquark, SO(10) grand unification, LHC} 

\maketitle

\section{Introduction} \label{sec:intro}

\noindent
During the past few years, several disagreements between experiments and the Standard Model (SM) predictions in the rare $B$ decays have been reported by the BaBar~\cite{Lees:2012xj,Lees:2013uzd}, 
LHCb~\cite{Aaij:2014ora,Aaij:2017vbb,Aaij:2015yra,Aaij:2017uff,Aaij:2017deq} and 
Belle~\cite{Huschle:2015rga,Sato:2016svk,Hirose:2016wfn,Hirose:2017dxl} collaborations. So far, these anomalies have been quite persistent. The most significant ones have been observed in the $R_{D^{(*)}}$ and $R_{K^{(*)}}$ observables, defined as,
\ba
R_{D^{(*)}}  &=&  \dfrac{\mathrm{BR}(B\rightarrow D^{(*)}\tau\nu)}{\mathrm{BR}(B\rightarrow D^{(*)}\ell\nu)} \quad \mbox{and} \nn\\ R_{K^{(*)}} &=& \dfrac{\mathrm{BR}(B\rightarrow K^{(*)}\mu^+\mu^-)}{\mathrm{BR}(B\rightarrow K^{(*)}e^+e^-)}\;.\nn
\ea
Here, $\ell=\{e$ or $\mu\}$ and $\mathrm{BR}$ stands for branching ratio. The experimental values of $R_D$ and $R_{D^*}$ are in excess of their SM predictions~\cite{Bigi:2016mdz,Bernlochner:2017jka,Bigi:2017jbd,Jaiswal:2017rve} by $1.4\sigma$ and $2.5\sigma$, respectively, (combined excess of $3.08\sigma$ in $R_{D^{(*)}}$) based on the world averages as of spring 2019, according to the Heavy Flavor Averaging Group~\cite{Amhis:2016xyh}, whereas, 
the observed $R_K$ and $R_{K^*}$ are both suppressed compared to their SM predictions~\cite{Hiller:2003js,Bordone:2016gaq} by $\sim 2.6\sigma$.

One of the possible explanations for the $B$-decay anomalies is the existence of scalar leptoquarks whose masses are 
in the few-TeV range
\cite{Dorsner:2013tla,
Sakaki:2013bfa,
Freytsis:2015qca,
Bauer:2015knc,
Dumont:2016xpj,
Das:2016vkr,
Becirevic:2016oho,
Becirevic:2016yqi,
Faroughy:2016osc,
Hiller:2016kry,
Chen:2017hir,
Crivellin:2017zlb,
Becirevic:2017jtw,
Cai:2017wry,
Altmannshofer:2017poe,
Assad:2017iib,
Jung:2018lfu,
Biswas:2018jun,
Bandyopadhyay:2018syt,
Aydemir:2018cbb,
Marzocca:2018wcf,
Becirevic:2018afm,
Kumar:2018kmr,
Hu:2018lmk,
Faber:2018qon,
Heeck:2018ntp,
Angelescu:2018tyl,
Bifani:2018zmi,
Bansal:2018nwp,
Mandal:2018kau,
Iguro:2018vqb,
Aebischer:2018acj,
Bar-Shalom:2018ure,
Kim:2018oih,
Arnan:2019olv})\footnote{See Refs.~\cite{Bhattacharya:2016mcc,Sahoo:2018ffv,Crivellin:2018yvo,Biswas:2018snp,Balaji:2018zna,Biswas:2018iak,Roy:2018nwc,Fornal:2018dqn} for the vector-leptoquark solutions proposed to explain the $B$-decay anomalies.}. Leptoquarks, which posses both lepton and quark couplings, often exist in the grand unified theories (GUTs) or in the Pati-Salam-type models. Considering that the LHC searches, except for these anomalies, have so far returned empty handed, if the leptoquarks indeed turn out to be behind these anomalies, it is likely that there will only be a small number of particles to be discovered. However, their existence in small numbers at the TeV scale would be curious in terms of its implications regarding physics beyond the Standard Model. Scalars in these models mostly come in large multiplets and it would be peculiar that only one or a few of the components become light at the TeV scale while others remain heavy. Mass splitting is actually a well-speculated subject in the literature in the context of the infamous doublet-triplet splitting problem in supersymmetric theories and GUTs. From this point of view, even the SM Higgs in a GUT framework is troublesome if it turns out to be the only scalar at the electroweak (EW) scale.

In this paper, we consider a single scalar leptoquark, $S_1(3,1,-1/3)$, at the TeV scale, in 
the $\mathrm{SO}(10)$ GUT framework~\cite{Fritzsch:1974nn,Georgi:1974my,Chang:1983fu,Chang:1984uy,Chang:1984qr,Deshpande:1992au,Bajc:2005zf,Bertolini:2009qj,Babu:2012vc,Altarelli:2013aqa,Aydemir:2015oob,Aydemir:2016qqj,Babu:2015bna,Babu:2016bmy}. This particular leptoquark was discussed in the literature to be responsible for one (or possibly both) of the $R_{D^{(*)}}$ 
and $R_{K^{(*)}}$ anomalies~\cite{Bauer:2015knc,Becirevic:2016oho,Cai:2017wry,Freytsis:2015qca}.
(Later, however, it was shown in Ref.~\cite{Angelescu:2018tyl} that a single $S_1$ leptoquark can only
alleviate the $R_{K^{(*)}}$ discrepancy, but cannot fully resolve it.)
Furthermore, it is contained in a relatively small multiplet that resides in the fundamental representation of the $\mathrm{SO}(10)$ group, a real $\mathbf{10}$, together with a scalar doublet with the quantum numbers that allow it to be identified as the SM Higgs. Therefore, $S_1$ being the only light scalar entity other than the SM Higgs doublet is justified in this scenario. A future discovery of such a leptoquark at the TeV scale could be interpreted as evidence in favour of an $\mathrm{SO}(10)$ GUT.

It has been argued in the literature that a real $\mathbf{10}_H$ in the minimal $\mathrm{SO}(10)$ setup (even together with a real $\mathbf{120}_H$ or a complex $\mathbf{120}_H$) is not favoured in terms of a realistic Yukawa sector~\cite{Bajc:2005zf}. On the other hand, it has been recently discussed in Ref.~\cite{Babu:2016bmy} that a Yukawa sector consisting of a real $\mathbf{10}_H$,  a real $\mathbf{120}_H$ and a complex $\mathbf{126}_H$, can establish a realistic Yukawa sector due to the contributions from the scalars whose quantum numbers are the same as the SM Higgs doublet. Thus, this is the scalar content we assign in our model for the Yukawa sector.

The inclusion of $S_1$ in the particle content of the model at the TeV scale does not improve the status of the SM in terms of gauge coupling unification, which cannot be realized by the particle content in question. Fortunately, in the $\mathrm{SO}(10)$ framework, there are other ways to unify the gauge coupling constants, in contrast to models based on the $\mathrm{SU}(5)$ group which also contains such a leptoquark within the same multiplet as the SM Higgs. As we illustrate in this paper, inserting a
single intermediate phase where the active gauge group is the Pati-Salam group, 
$\mathrm{SU}(4)_{\mathcal{C}}\otimes \mathrm{SU}(2)_L\otimes \mathrm{SU}(2)_R$, which appears to be the favoured route of symmetry-breaking by various phenomenological bounds~\cite{Altarelli:2013aqa}, establishes coupling unification as desired. In our model, the Pati-Salam group is broken into the SM gauge group at an intermediate energy scale $M_C$, while $\mathrm{SO}(10)$ is broken into the Pati-Salam group at the unification scale $M_U$. We consider two versions of this scenario, depending on whether the left-right symmetry, so-called $D$ parity, is broken together with $\mathrm{SO}(10)$ at $M_U$, or it is broken at a later stage, at $M_C$, where the Pati-Salam symmetry is broken into the SM gauge symmetry.

Light colour triplets, similar to the one we consider in this paper, are often dismissed for the sake of proton stability since these particles in general have the right quantum numbers for them to couple potentially dangerous operators that mediate proton decay. On the other hand, the proton stability could possibly be ensured through various symmetry mechanisms such as the utilization of Peccei-Quinn (PQ) symmetry~\cite{Cox:2016epl,Bajc:2005zf}, other $\mathrm{U}(1)$ symmetries such as the one discussed in Ref.~\cite{Pati:1974yy}, or a discrete symmetry similar to the one considered in Ref.~\cite{Bauer:2015knc}. Operators leading to proton decay could also be suppressed by a specific mechanism such as the one discussed in Ref.~\cite{Dvali:1995hp} or they could be completely forbidden by geometrical reasons~\cite{Aydemir:2018cbb}. In this paper, we adopt a discrete symmetry as suggested in Ref.~~\cite{Bauer:2015knc}, assumed to operate below the intermediate symmetry-breaking scale even though it is not manifest at higher energies.

Motivated by the possible existence of a single TeV scale $S_1$ in the $\mathrm{SO}(10)$ GUT 
framework, we move on to investigate the phenomenological implications of our model. In Ref.~\cite{Bauer:2015knc}, it was shown that a TeV scale $S_1$ leptoquark can explain the $R_{D^{(*)}}$ anomalies while simultaneously inducing the desired suppression in $R_{K^{(*)}}$ through box diagrams. Since the most significant anomalies are
seen in the $R_{D^{(*)}}$ observables, in this paper, we concentrate mostly on scenarios that can accommodate these observables. 
Generally, a TeV scale $S_1$ requires 
one Yukawa coupling to be large to accommodate the  $R_{D^{(*)}}$ anomalies~\cite{Cai:2017wry,Angelescu:2018tyl}. This, 
however, could create a problem for the $b\to s\bar{\nu}\nu$ transition rate measured in the $R_{K}^{\nu\nu}$ observable.
In the SM, this decay  proceeds through a loop
whereas $S_1$ can contribute at the tree level in this transition. Therefore, the measurement of 
$R_{K}^{\nu\nu}$ is very important to restrict the parameter space of $S_1$.\footnote{One can avoid this conflict by introducing some additional degrees of freedom, as shown in Ref.~\cite{Crivellin:2017zlb}. There, the authors introduced an $S_3$ leptoquark in 
addition to the $S_1$ to concomitantly explain $R_{D^{(*)}}$ and $R_{K^{(*)}}$ while being consistent with $R_{K}^{\nu\nu}$.} 
Some specific Yukawa couplings of $S_1$ are also severely constrained from the $Z\to \tau\tau$ decay~\cite{Bansal:2018nwp} and the 
LHC $\tau\tau$ resonance search data~\cite{Mandal:2018kau}.
Therefore, it is evident that in order to find the $R_{D^{(*)}}$-favoured parameter space while successfully accommodating other 
relevant constraints, one has to introduce new degrees of freedom
in terms of new couplings and/or new particles.
Here, we consider a specific Yukawa texture with three free couplings to show that a TeV scale 
$S_1$, consistent with relevant measurements, can still explain the $R_{D^{(*)}}$ anomalies.
   
The rest of the paper is organized as follows. In Sec. \ref{sec:model}, we introduce our model. In Sec. \ref{sec:unification}, we display the unification of the couplings for two versions of our model. In Sec. \ref{sec:pheno}, we present the related LHC phenomenology with a single extra leptoquark $S_1$. We display the exclusion limits from the LHC data and discuss related future prospects. We also study the renormalization group (RG) running of the Yukawa couplings. Finally in Sec. \ref{sec:outlook}, we end our paper with a discussion and conclusions.

\section{The $\mathbf{SO(10)}$ model}
\label{sec:model}
\noindent
In our $\mathrm{SO}(10)$ model, we entertain the idea that the SM Higgs doublet is not the only scalar multiplet at the TeV scale, but it is accompanied by a leptoquark $S_1=(3,1,-1/3)$, both of which reside in a real ten-dimensional representation, $\mathbf{10}$, of $\mathrm{SO}(10)$ group. The peculiar mass splitting among the components of this multiplet does not occur, leading to a naturally light scalar leptoquark at the TeV scale. 

We start with a real  $\mathbf{10}_H$ of $\mathrm{SO}(10)$ whose Pati-Salam and SM decompositions are given as
\begin{eqnarray}
\mathbf{10}&=& \left(1,2,2\right)_{422} \oplus \left(6,1,1\right)_{422} \nonumber\\ \nonumber\\
&=& \underbrace{\left(1,2,\frac{1}{2}\right)_{321}}_{H}\oplus \underbrace{ \left(1,2,-\frac{1}{2}\right)_{321}}_{H^*}\oplus\underbrace{ \left(3,1,-\frac{1}{3}\right)_{321}}_{S_1}\nn\\&&\oplus \underbrace{\left(\overline{3},1,\frac{1}{3}\right)_{321}}_{S_1^{*}},
\end{eqnarray}
where subscripts denote the corresponding gauge group and we set $Q=I_3+Y$.

The scalar content we assign for the Yukawa sector consists of a real $\mathbf{10}_H$, a real $\mathbf{120}_H$, and a complex $\mathbf{126}_H$, which establishes a realistic Yukawa sector through mixing between the scalars whose quantum numbers are the same as the SM Higgs doublet, as shown in Ref.~\cite{Babu:2016bmy}.
Note that
\begin{eqnarray}
\mathbf{16}\otimes \mathbf{16}=\mathbf{10}_s\oplus \mathbf{120}_a\oplus\mathbf{126}_s,
\end{eqnarray}
where $\mathbf{16}$ is the spinor representation in which each family of fermions, including the right-handed neutrino, resides in. The subscripts $s$ and $a$ denote the symmetric and antisymmetric components. The Pati-Salam decompositions of $\mathbf{120}$ and $\mathbf{126}$ are given as
\begin{eqnarray}
\mathbf{120}&=&\left(1,2,2\right)_{422}\oplus \left(1,1,10\right)_{422}\oplus \left(1,1,\overline{10}\right)_{422}  \nn\\
&&\oplus \left(6,3,1\right)_{422}  \oplus \left(6,1,3\right)_{422} \oplus  \left(15,2,2\right)_{422},                  \nonumber\\ \mathbf{126}&=&\left(10,3,1\right)_{422}\oplus \left(\overline{10},1,3\right)_{422} \oplus \left(15,2,2\right)_{422} \nn\\
&&\oplus \left(6,1,1\right)_{422}\;. 
\end{eqnarray}
The Yukawa terms are then given as 
\begin{eqnarray}
\mathcal{L}_Y&=&\mathbf{16}_F\left(Y_{10}\mathbf{10}_H+Y_{120}\mathbf{120}_H+Y_{126}\mathbf{\overline{126}}_H\right)\mathbf{16}_F\;\nn\\
&&+ \mathrm{H.c.},
\end{eqnarray}
where $Y_{10}$ and $Y_{126}$ are complex Yukawa matrices, symmetric in the generation space, and $Y_{120}$ is a complex antisymmetric one.

The SM doublet contained in $\phi(1,2,2)_{422}$ of $\mathbf{10}$ is mixed with other doublets accommodated  in $\phi(1,2,2)_{422}$ and $\Sigma(15,2,2)_{422}$ of the real multiplet $\mathbf{120}$ and  $\Sigma(15,2,2)_{422}$ of  $\mathbf{126}$, yielding a Yukawa sector consistent with the observed fermion masses~\cite{Babu:2016bmy}. The fermion mass matrices for the up quark, down quark, charged leptons, Dirac neutrinos and Majorana neutrinos are given as~\cite{Babu:2016bmy}
\begin{eqnarray}
\label{massmatrices}
M_U&=& v_{10} Y_{10}+v^{u}_{126} Y_{126}+(v^{(1)}_{120}+v_{120}^{(15)})Y_{120}\;,\nonumber\\
M_D&=& v_{10} Y_{10}+v^{d}_{126} Y_{126}+(v^{(1)}_{120}+v_{120}^{(15)})Y_{120}\;,\nonumber\\
M_E&=& v_{10} Y_{10}-3v^{d}_{126} Y_{126}+(v^{(1)}_{120}-3v_{120}^{(15)})Y_{120}\;,\nonumber\\
M_{\nu_D}&=& v_{10} Y_{10}-3v^{u}_{126} Y_{126}+(v^{(1)}_{120}-3v_{120}^{(15)})Y_{120}\;,\nonumber\\
M_{\nu_{R,L}}&= &v_{R,L} Y_{126}\;,
\end{eqnarray}
where $v^{u}_{i}$($v^{d}_{i}$) are the presumed vacuum expectation values of the neutral components of the corresponding bidoublets, contributing to the masses of the up quarks and the Dirac neutrinos (the down quarks and the charged leptons), and where
\begin{eqnarray}
\label{realitycondition}
v_{10}&\equiv& v^{u}_{10}=v^{d\;*}_{10}\;,\nonumber\\
v_{120}^{(1)}&\equiv& v^{u, (1)}_{120}=v^{d,(1)\;*}_{120}\;,\nonumber\\
v_{120}^{(15)}&\equiv& v^{u, (15)}_{120}=v^{d,(15)\;*}_{120}\;,
\end{eqnarray}
due to the reality condition of $\mathbf{10}$ and $\mathbf{120}$). The superscripts $(1)$ and $(15)$ denote the doublets residing in  $\phi(1,2,2)_{422}$ and $\Sigma(15,2,2)_{422}$ of $\mathbf{120}$, respectively. $v_{R}$ in the Majorana mass matrices given in Eq.~(\ref{massmatrices}) is the vacuum expectation value of the SM singlet contained in  $(\overline{10},1,3)_{422}$ and is responsible for the heavy masses of the right-handed neutrinos (type-I seesaw), and $v_{L}\sim v_{R} v_{10}^2/M^2_{GUT}$ is induced by the potential term  $\mathbf{10}^2_H \mathbf{120}^2_H$ and is responsible for the left-handed Majorana neutrino masses (type-II seesaw)~\cite{Babu:2016bmy}.

For the symmetry-breaking pattern, we consider a scenario in which the intermediate phase has the gauge symmetry of the Pati-Salam group $\mathrm{SU}(4)_{\mathcal{C}}\otimes \mathrm{SU}(2)_L\otimes \mathrm{SU}(2)_R$.  Note that this specific route of symmetry-breaking appears to be favoured by
various phenomenological bounds~\cite{Altarelli:2013aqa}. We consider two versions of this scenario depending on whether or not the Pati-Salam gauge symmetry is accompanied  by the $D$-parity invariance, a $\mathbb{Z}_2$ symmetry that maintains the complete equivalence of the left and right sectors~\cite{Chang:1984uy,Chang:1984qr,Maiezza:2010ic}, after the $\mathrm{SO}(10)$ breaking. The symmetry-breaking sequence is schematically given as
\begin{widetext}
\begin{eqnarray}
\label{chain}
 \quad \mathrm{SO}(10) \;\underset{\left<\mathbf{ 210}\right> \;(\mbox{\tiny or } \left<\mathbf{ 54}\right>)}{\xrightarrow{M_U}}\; G_{422\;(\mbox{\tiny or}\;422D)} \;\underset{\left<\mathbf{126}\right>}{\xrightarrow{M_C}}\; G_{321}\;\mbox{(SM)} \;\underset{\left<\mathbf{10}\right>}{\xrightarrow{M_Z}}\; G_{31},
\end{eqnarray}
\end{widetext}
%
where, we use the notation,
\begin{eqnarray}
G_{422D}&\equiv& \mathrm{SU}(4)_{\mathcal{C}}\otimes \mathrm{SU}(2)_L\otimes \mathrm{SU}(2)_R\otimes  D,\cr
G_{422}&\equiv& \mathrm{SU}(4)_{\mathcal{C}}\otimes \mathrm{SU}(2)_L\otimes \mathrm{SU}(2)_R,\cr
G_{321}&\equiv& \mathrm{SU}(3)_C\otimes \mathrm{SU}(2)_L\otimes \mathrm{U}(1)_{Y},\cr
G_{31}&\equiv& \mathrm{SU}(3)_C\otimes \mathrm{U}(1)_{Q}\;.
\end{eqnarray}
The first stage of the spontaneous symmetry-breaking occurs through the Pati-Salam singlet in the 
$\mathrm{SO}(10)$ multiplet $\mathbf{210}$, acquiring a vacuum expectation value (VEV) at the unification scale $M_U$.  This singlet is odd under $D$ parity and, therefore, the resulting symmetry group is $G_{422}$ in the first stage of the symmetry-breaking. In the second step, the breaking of $G_{422}$ into the SM gauge group $G_{321}$ is realized through the SM singlet contained in $\Delta_R(\overline{10},1,3)_{422}$ of $\mathbf{126}$, acquiring VEV at the energy scale $M_C$, which also yields a Majorana mass for the right-handed neutrino. The last stage of the symmetry-breaking is realized predominantly through the SM doublet contained in $\phi(1,2,2)_{422}$ of $\mathbf{10}$. The mass scale of these Pati-Salam multiplets is set as $M_C$, the energy scale at which the Pati-Salam symmetry is broken into the SM, while the rest of the fields are assumed to be heavy at the unification scale $M_U$. The only degrees of freedom, assumed to survive down to the electroweak scale, are the SM Higgs doublet and the colour triplet, $S_1$. We call this model $A_1$. 

In the second scenario, which we call model $A_2$, the first stage of the symmetry-breaking is realized through the Pati-Salam singlet contained in $\mathbf{54}$, which acquires a VEV. This singlet is even under $D$ parity, and therefore, $D$ parity is not broken at this stage with the $\mathrm{SO}(10)$ symmetry, and the resulting symmetry group valid down to $M_C$ is $G_{422}$. The rest of the symmetry-breaking continues in the same way as in model $A_1$. Consequently in model $A_2$, we include one more Pati-Salam multiplet at $M_C$, $\Delta_L(10,3,1)$, in order to maintain a complete left-right symmetry down to $M_C$. The scalar content and, for later use, the corresponding RG coefficients in each energy interval are given in Table~\ref{a1B}.

Finally, the relevant Lagrangian for phenomenological analysis at low energy is given by
\begin{eqnarray}
\label{eq:lagrangianLQ}
\mathcal{L}&\supset&  \left(D_{\mu} S_1\right)^{\dag} \left(D^{\mu} S_1\right)-M_{S_1}^2 |S_1|^2-\lambda |S_1|^2 |H|^2\nn\\
&&+\lt(\boldsymbol{\Lambda}^{L}\bar{Q}^c i\tau_2 L+\boldsymbol{\Lambda}^{R}\bar{u}_R^c e_R\rt)S_1^{\dag} +\mathrm{H.c.},
\end{eqnarray}
where $Q$ and $L$ are the SM quark and lepton doublets (for each family), 
$\boldsymbol{\Lambda}^{L/R}$ are coupling matrices in flavour space and $\psi^c=C\bar{\psi}^T$ are charge-conjugate spinors. Notice the absence of dangerous diquark couplings of $S_1$ that would lead to proton decay. One way to forbid these couplings is to impose a $\mathbb{Z}_2$ symmetry~\cite{Bauer:2015knc} that emerges below the Pati-Salam breaking scale under which quarks and leptons transform with opposite parities whereas the leptoquark is assigned odd parity, i.e.~$(q, l, S_1)\rightarrow (\pm q, \mp l, -S_1)$. Note also that the inclusion of $S_1$ can affect the stability of the electroweak vacuum via loop
effects. The relevant discussion can be found in Ref.~\cite{Bandyopadhyay:2016oif}.

Evidently, the Lagrangian given in Eq.~(\ref{eq:lagrangianLQ}) together with the SM Lagrangian should be understood in the effective field theory context. The new and SM Yukawa couplings in the TeV scale Lagrangian are induced from the original $\mathrm{SO}(10)$ Yukawa couplings each of which is generated by a linear combination of unification-scale operators and gets modified due to the mixing effects induced by the scalar fields that have Yukawa couplings to three chiral families of  $\mathbf{16}_F$.  It is indeed this rich structure that enables the realization of a fermion mass spectrum consistent with the expected fermion masses of the SM model at the unification scale, as shown in Ref.~\cite{Babu:2016bmy}. As we will discuss later in Sec.~\ref{sec:Yukawa}, the modification to the SM RG running of the Yukawa couplings due to the inclusion of $\boldsymbol{\Lambda}^{L/R}$ does not register strong changes in the fermionic mass spectrum and hence the main message of Ref.~\cite{Babu:2016bmy} is valid in our case, as well.

\begin{table*}
\caption{The scalar content and the RG coefficients in the energy intervals for model $A_{1,2}$. Note that the $\phi$ fields,  the $\Phi$ field, and one of the $\Sigma$ fields originate from real 
$\mathrm{SO}(10)$ multiplets and thus the $\eta=1/2$ condition should be employed when necessary while determining the RG coefficients in Eq.~\eqref{1loopgeneral}.}
\begin{center}
{\begin{tabular}{c|c|c}
\hline
$\vphantom{\Big|}$ Interval & Scalar content for model $A_1$ ($A_2$) & RG coefficients
\\
\hline\hline
$\vphantom{\Biggl|}$ II
&  $\phi(1,2,2)\times 2$, $\Phi(6,1,1)$,\vspace{-0.05cm}&\vspace{-0.2cm}
\\
& $\Sigma(15,2,2)\times 2$, $\Delta_R(\overline{10},1,3)$, &$\left[ a_{4},a_{L},a_{R}\right]=\left[\dfrac{1}{2}\left(\dfrac{7}{2}\right),\dfrac{9}{2}\left(\dfrac{67}{6}\right),\dfrac{67}{6}\right]$\\
&$\left(\mbox{and } \Delta_L(10,3,1)\mbox{ for model } A_2\right)$
& $\vphantom{\Big|}$ 
\\
\hline
$\vphantom{\Biggl|}$   I
& $H\left(1,2,\dfrac{1}{2}\right)$, $S_1\left(3,1,-\dfrac{1}{3}\right)$
& $\left[a_{3},a_{2},a_{1}\right]=\left[\dfrac{-41}{6},\dfrac{-19}{6},\dfrac{125}{18}\right]$
\\
\hline
\end{tabular}}
\label{a1B}
\end{center}
\end{table*}

\section{Gauge coupling unification \label{sec:unification}}
\noindent
In this section, after we lay out the preliminaries for one-loop RG running and show that the new particle content at the TeV scale does not lead to the unification of the SM gauge couplings directly, we illustrate gauge coupling unification with a single intermediate step of symmetry-breaking. Once the particle content at low energies is determined, there may be numerous ways to unify the gauge couplings, depending on the selection of the scalar content in $\mathrm{SO}(10)$ representations. In the literature, the canonical way to make this selection is through adopting a minimalistic approach, allowed by the observational constraints. In the following, we pursue the same strategy while taking into account the analysis made in Ref.~\cite{Babu:2016bmy} for a realistic Yukawa sector. 

\subsection{One-loop RG running}
\noindent
For a given particle content, the gauge couplings in an energy interval $\left[M_A,M_B\right]$ evolve under one-loop RG running as
\begin{eqnarray}
\label{gaugerunning}
\frac{1}{g_{i}^{2}(M_A)} - \dfrac{1}{g_{i}^2(M_B)}
\;=\; \dfrac{a_i}{8 \pi^2}\ln\dfrac{M_B}{M_A},
\end{eqnarray}
where the RG coefficients $a_i$ are given by \cite{Jones:1981we,Lindner:1996tf}
\begin{eqnarray}
\label{1loopgeneral}
a_{i}
\;=\; -\frac{11}{3}C_{2}(G_i)
& + & \frac{2}{3}\sum_{R_f} T_i(R_f)\cdot d_1(R_f)\cdots d_n(R_f) \cr
& + & \frac{\eta}{3}\sum_{R_s} T_i(R_s)\cdot d_1(R_s)\cdots d_n(R_s),\nn\\
\end{eqnarray}
and the full gauge group is given as $G=G_i\otimes G_1\otimes...\otimes G_n$.
The summation in Eq.~\eqref{1loopgeneral} is over irreducible chiral representations of fermions ($R_f$) and irreducible representations of scalars ($R_s$) in the second and third terms, respectively. The coefficient $\eta$ is either 1 or 1/2, depending on whether the corresponding representation is complex or (pseudo) real, respectively. $d_j(R)$ is the dimension of the representation $R$ under the group $G_{j\neq i}$.~$C_2(G_i)$ is the quadratic Casimir for the adjoint representation of the group $G_i$,
and $T_i$ is the Dynkin index of each representation (see Table~\ref{DynkinIndex}). For $\mathrm{U}(1)$ group, $C_2(G)=0$ and
\begin{equation}
\sum_{f,s}T \;=\; \sum_{f,s}Y^2,
\label{U1Dynkin}
\end{equation}
where $Y$ is the $\mathrm{U}(1)_Y$ charge.

\begin{figure*}[t!]
\captionsetup[subfigure]{labelformat=empty}
\subfloat[\quad\quad\quad(a)]{\includegraphics[width=7cm]{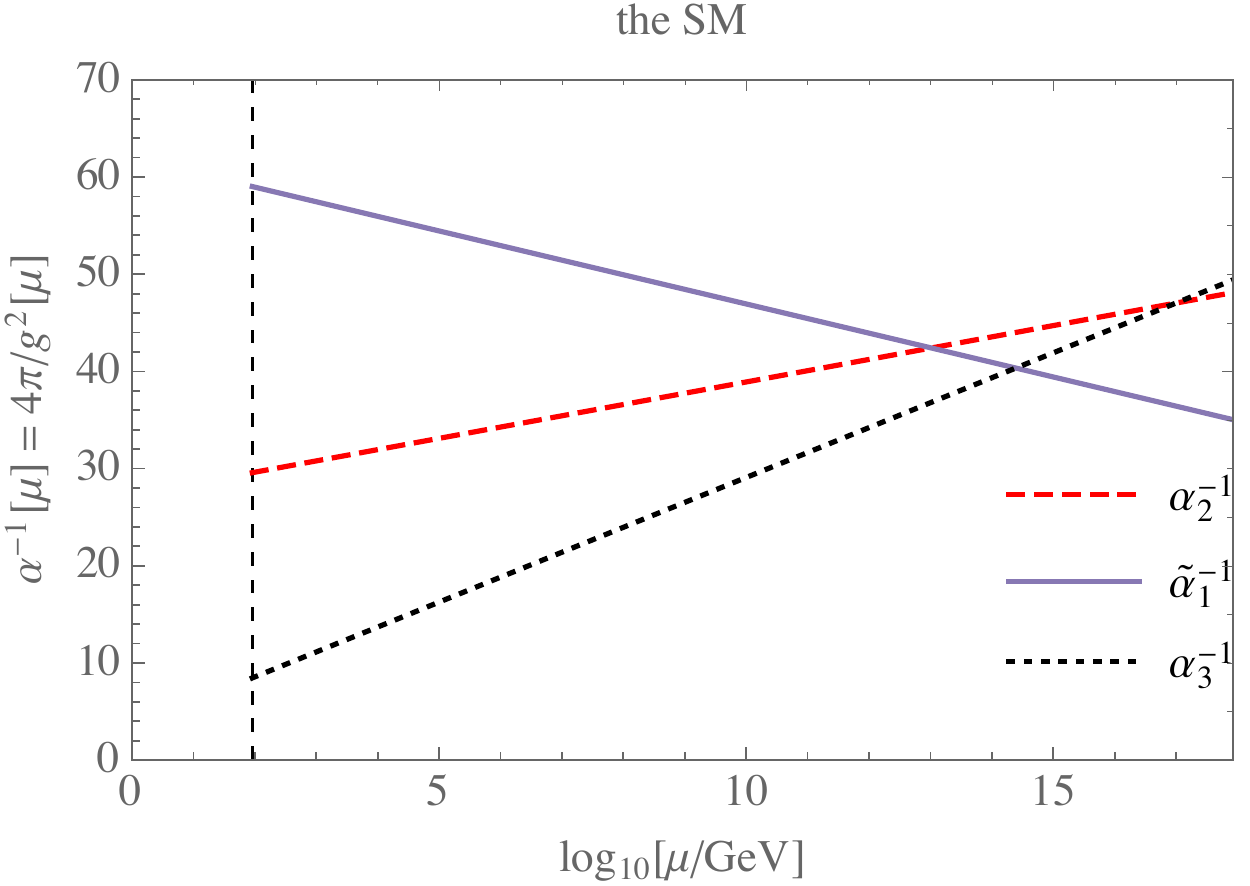}}\hspace{0.4cm}
\subfloat[\quad\quad\quad(b)]{\includegraphics[width=7cm]{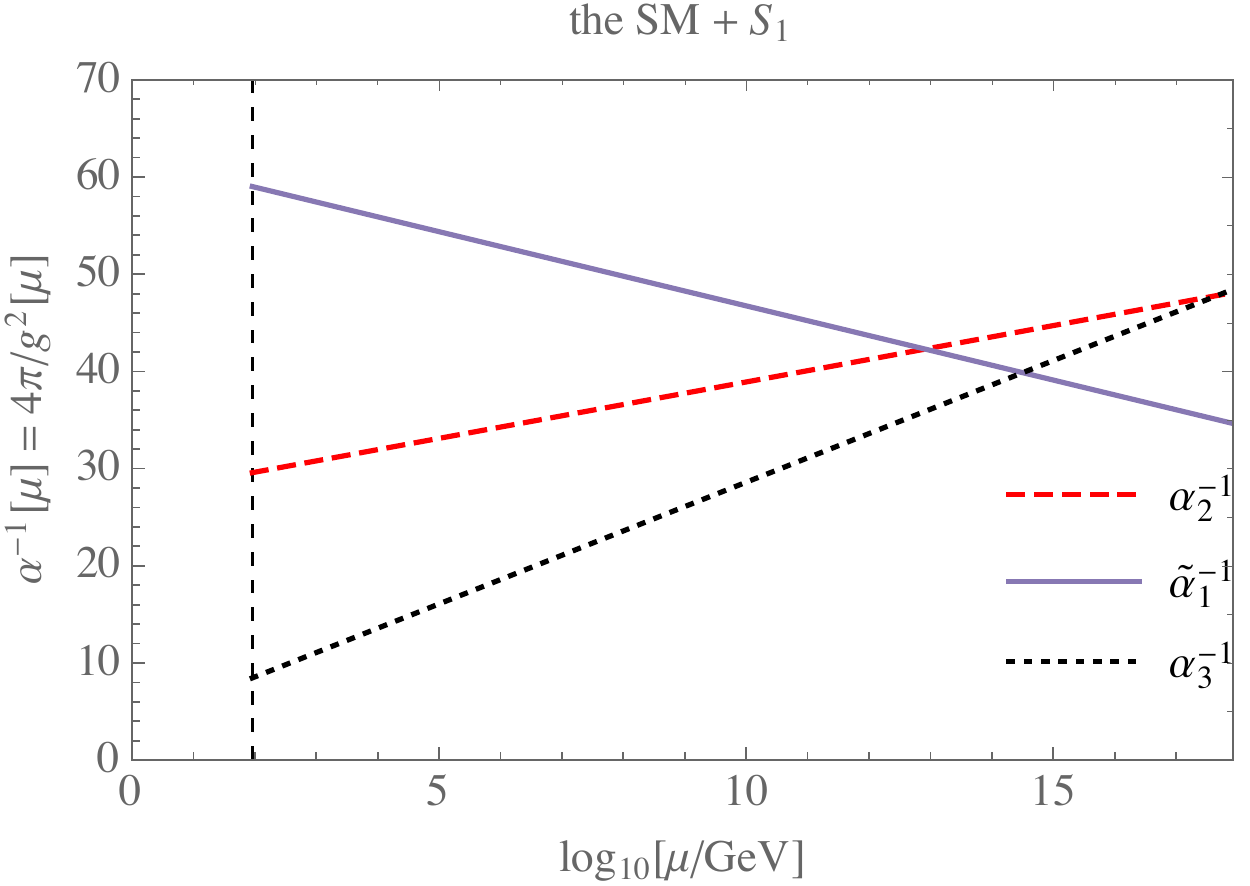}}\hspace{0.4cm}
\caption{Running of the gauge couplings with the particle content of the SM and with the inclusion of $S_1$ . The vertical dotted line corresponds to the electroweak scale $M_Z$. For $\alpha_1^{-1}$, we plot the redefined quantity $\widetilde{\alpha}^{-1}_1\equiv \frac{3}{5}\alpha^{-1}_1$ as required by the $\mathrm{SO}(10)$ boundary conditions. Including the leptoquark in the particle content does not provide a significant modification to the SM RG running in favour of unification.}
\label{RGrunning1}
\end{figure*}

The addition of $S_1$ to the particle content of the SM does not help in unifying the gauge couplings as displayed in Fig.~\ref{RGrunning1}, where the RG running is performed with the modified RG coefficients given in Table~\ref{a1B} in interval I, while interval II is irrelevant to this particular case. Unification of the gauge couplings can be established through intermediate symmetry-breaking between the electroweak scale and unification scale, as we illustrate in the next subsection with a single intermediate step of symmetry-breaking.

\begin{table}[t!]
\caption{Dynkin index $T_i$ for various irreducible representations of $\mathrm{SU}(2)$, $\mathrm{SU}(3)$, and $\mathrm{SU}(4)$. 
Our normalization convention in this paper follows the one adopted in Ref.~\cite{Lindner:1996tf}. Notice that there are two inequivalent 15-dimensional irreducible representations for 
$\mathrm{SU}(3)$.}
\begin{center}
{\begin{tabular}{ccccc}
\hline
\ \ \ Representation & $\quad \mathrm{SU}(2)\quad$ & $\quad \mathrm{SU}(3)\quad$ & $\quad \mathrm{SU}(4)\quad$ &\\ 
\hline\hline
$\vphantom{\bigg|}$ 2 &   $\dfrac{1}{2}$ &              $-$ &   $-$ & $\vphantom{\bigg|}$ \\
$\vphantom{\bigg|}$ 3 &                2 &   $\dfrac{1}{2}$ &   $-$ & $\vphantom{\bigg|}$ \\
$\vphantom{\bigg|}$ 4 &                5 &              $-$ &   $\dfrac{1}{2}$ & $\phantom{\bigg|}$ \\ 
$\vphantom{\bigg|}$ 6 &  $\dfrac{35}{2}$ &   $\dfrac{5}{2}$ &   $1$ & $\vphantom{\bigg|}$ \\
$\vphantom{\bigg|}$ 8 &               42 &              $3$ &   $-$ & $\vphantom{\bigg|}$ \\
$\vphantom{\bigg|}$10 & $\dfrac{165}{2}$ &  $\dfrac{15}{2}$ &   $3$ & $\vphantom{\bigg|}$ \\
$\vphantom{\bigg|}$15 &              280 & $10,\dfrac{35}{2}$ &  4 & $\vphantom{\bigg|}$ \\
\hline
\end{tabular}
\label{DynkinIndex}}
\end{center}
\end{table}

\subsection{Unification with a single intermediate scale} 
\noindent
We start by labelling the energy intervals in between symmetry-breaking scales
$[M_Z,M_C]$ and $[M_C,M_U]$ with Roman numerals as
\begin{eqnarray}
\mathrm{I}  & \;:\; & [M_Z,\;M_C],\quad G_{213}  \;(\mathrm{SM}) ,\cr
\mathrm{II} & \;:\; & [M_C,\;M_U],\quad G_{224}\quad\mbox{or}\quad G_{224D} \;.
\label{IntervalNumber}
\end{eqnarray}
The boundary/matching conditions we impose on the couplings at the symmetry-breaking scales are
\begin{eqnarray}
M_U & \;:\; & g_L(M_U) \;=\; g_R(M_U) \;=\; g_4(M_U) , \vphantom{\bigg|} 
\cr
M_C & \;:\; & g_3(M_C)=g_4(M_C) ,\quad g_2(M_C)\;=\;g_L(M_C),\cr
 & \;\; & \frac{1}{g_1^2(M_C)} \;=\; \frac{1}{g_R^2(M_C)}+ \frac{2}{3}\frac{1}{g_4^2(M_C)},\nn\\
  & \;\; & g_L(M_C)\;=\;g_R(M_C) \mbox{ (absent in the $G_{224}$ case)}
\vphantom{\Bigg|}
\cr
M_Z & \;:\; & \frac{1}{e^2(M_Z)} \;=\; \frac{1}{g_1^2(M_Z)}+\frac{1}{g_2^2(M_Z)}\;.\vphantom{\Bigg|} 
\label{Matching}
\end{eqnarray}
We use the central values of the low-energy data as the boundary conditions in the RG running 
(in the $\overline{\mathrm{MS}}$ scheme)
\cite{Patrignani:2016xqp,ALEPH:2005ab}
$\alpha^{-1} = 127.95$, $\alpha_s = 0.118$, $\sin^2\theta_W = 0.2312$ at $M_Z=91.2\,\mathrm{GeV}$, which translate to $g_1 = 0.357$, $g_2 = 0.652$, $g_3 = 1.219$. 
The coupling constants are all required to remain in the perturbative regime during the
evolution from $M_Z$ to $M_U$.

\begin{figure*}[t!]
\captionsetup[subfigure]{labelformat=empty}
\subfloat[\quad\quad\quad(a)]{\includegraphics[width=7cm]{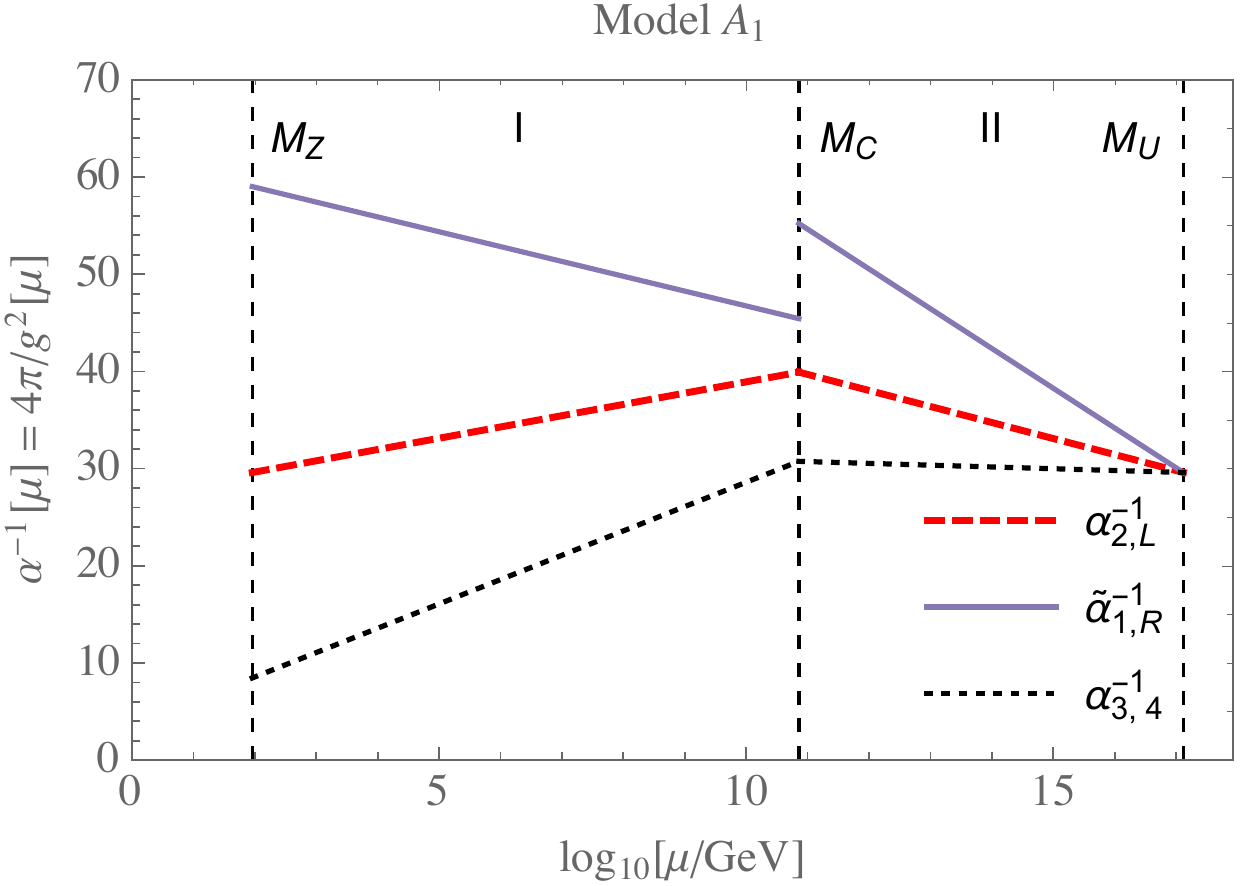}}\hspace{0.4cm}
\subfloat[ \quad\quad\quad(b)]{\includegraphics[width=7cm]{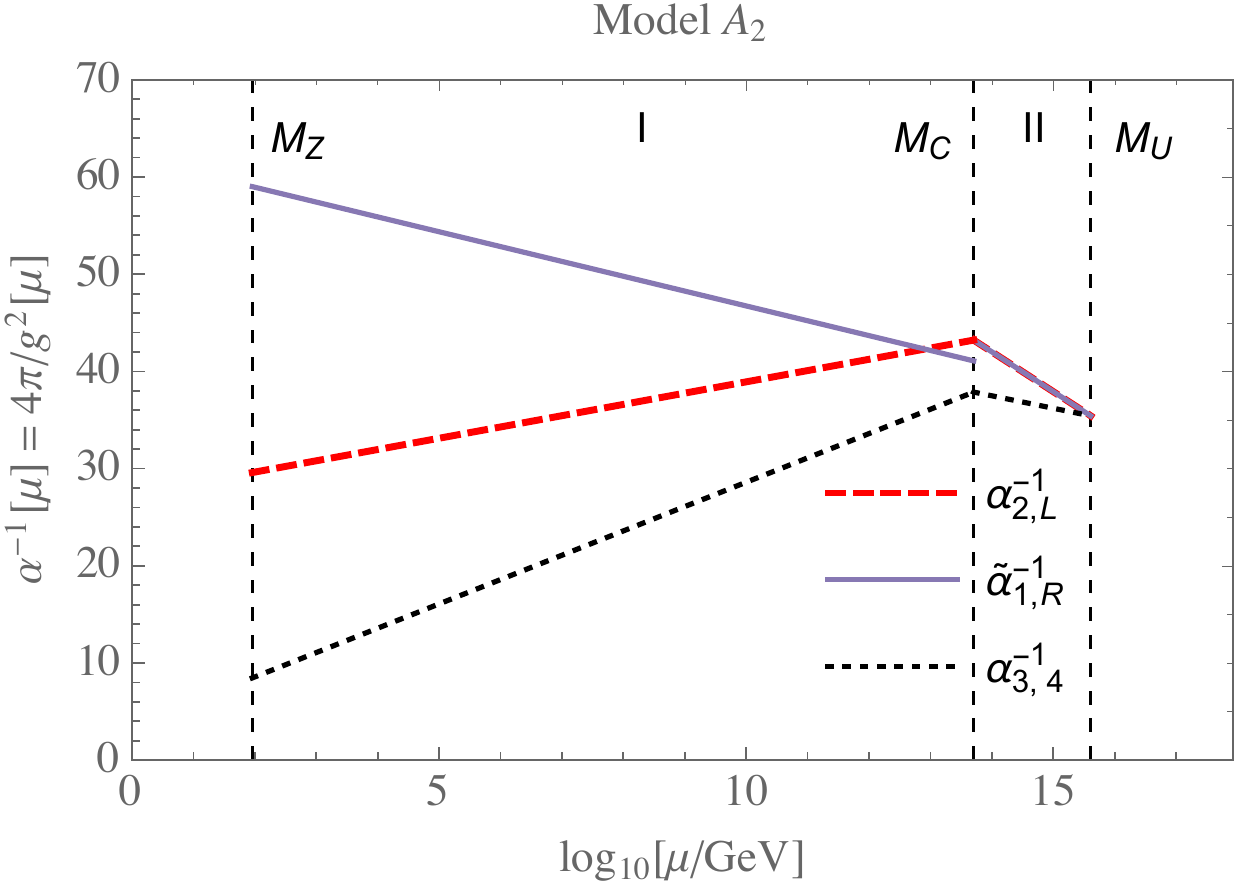}}\hspace{0.4cm}
\caption{Running of the gauge couplings for models $A_1$ and $A_2$. Note that $\widetilde{\alpha}^{-1}_1\equiv \frac{3}{5}\alpha^{-1}_1$. The discontinuity at $M_C$ in each plot is due to the boundary conditions given in Eq.~\eqref{Matching}.}
\label{RGrunning2}
\end{figure*}

The RG coefficients, $a_i$, differ depending on the particle content in each energy interval, changing every time symmetry-breaking occurs. Together with the matching and boundary conditions, 
one-loop RG running leads to the following conditions on the symmetry-breaking scales $M_U$ and $M_C$:
\begin{eqnarray}
2\pi\left[\dfrac{3-8\sin^2\theta_W(M_Z)}{\alpha(M_Z)}\right]
 = 
 \left(3a_1 -5a_2\right)\ln\dfrac{M_C}{M_Z}\hspace{1cm}&&\nn\\
\hspace{-.75cm}+\left(-5a_L+3a_R+2a_4\right)\ln\dfrac{M_U}{M_C}
,&&
\vphantom{\Bigg|}
\cr
2\pi\left[\dfrac{3}{\alpha(M_Z)} - \dfrac{8}{\alpha_s(M_Z)}\right]
 = 
 \left(3a_1 + 3a_2 - 8a_3\right)\ln\dfrac{M_C}{M_Z}&&\nn\\
\hspace{-0.75cm}+\left(3a_L+3a_R-6a_4\right)\ln\dfrac{M_U}{M_C}
,&&
\vphantom{\Bigg|}
\cr
& &\hspace{-0.75cm}~
\end{eqnarray}
where the notation on $a_i$ is self-evident.
The unified gauge coupling $\alpha_U$ at the scale $M_U$ is then obtained from
\begin{eqnarray}
\label{A6}
\dfrac{2\pi}{\alpha_U}
 =  \dfrac{2\pi}{\alpha_s(M_Z)}
-\left( a_4\;\ln\dfrac{M_U}{M_C}
+ a_3\;\ln\dfrac{M_C}{M_Z}
\right)
\;.\qquad
\end{eqnarray}
Thus, once the RG coefficients in each interval are specified,
the scales $M_U$ and $M_C$, and the value of $\alpha_U$ are uniquely determined. The results are given in Table~\ref{Results}, and unification of the couplings is displayed in Fig.~\ref{RGrunning2} for each model. 

As mentioned previously, we assume in this paper that the proton-decay-mediating couplings of $S_1$ are suppressed. On the other hand, we do not make any assumptions regarding the other potentially dangerous operators which could lead to proton decay. Thus, it is necessary to inspect whether the predictions of our models displayed in Table~\ref{Results} are compatible with the current bounds coming from the proton decay searches or not. The most recent and stringent bound on the lifetime of the proton comes from the mode $p\rightarrow e^+\pi^0$, and is $\tau_p >1.6\times 10^{34}$ years~\cite{Miura:2016krn}. As for the proton decay modes that are mediated by the super-heavy gauge bosons, which reside in the adjoint representation of $\mathrm{SO}(10)$ $\mathbf{45}$, considering that 
$\tau_p\sim M_U^4/m_p^5 \alpha_U^2$~\cite{Langacker:1980js}, we obtain $M_U\gtrsim 10^{15.9}$ GeV, which is consistent with predictions of both model $A_1$ and model $A_2$, within an order of magnitude of the latter. Additionally, since $M_C$ is the scale at which the Pati-Salam symmetry breaks into the SM, it determines the expected mass values for the proton-decay-mediating colour triplets. From a naive analysis~\cite{Altarelli:2013aqa}, it can be shown that the current bounds on the proton lifetime require $M_C\gtrsim 10^{11}$~GeV, again consistent with the predictions of both model $A_1$ and model $A_2$, within an order of magnitude of the former. Note that these bounds should be taken as order-of-magnitude estimates since, while obtaining them, we approximate the anticipated masses of the super-heavy gauge bosons and the colour triplets as $M_X\approx M_U$ and $M_T\approx M_C$, while it would not be unreasonable to expect that these mass values could differ from the corresponding energy scales within an order of magnitude.

\begin{table}[b!]
\caption{The predictions of models $A_1$ and $A_2$.}
\begin{center}
{\begin{tabular}{c||c|c}
\hline
$\vphantom{\Big|}$
\hspace{1cm}Model \hspace{1cm}
& \hspace{0.75cm} $A_1$ \hspace{0.75cm} 
& \hspace{0.75cm} $A_2$ \hspace{0.75cm} 
\\
\hline\hline
$\vphantom{\bigg|}$ $\log_{10}(M_U/\mathrm{GeV})$  &$17.1$& $15.6$ \\
$\vphantom{\bigg|}$ $\log_{10}(M_C/\mathrm{GeV})$ & $10.9$& $13.7$ \\ 
\hline
$\vphantom{\bigg|}$ $\alpha_U^{-1}$  & $29.6$& $35.4$ \\
\hline
\end{tabular}}
\label{Results}
\end{center}
\end{table}

\section{Low-energy phenomenology}
\label{sec:pheno}
\noindent
The existence of a TeV scale charge $-1/3$ scalar leptoquark in a GUT framework is quite interesting from a
phenomenological perspective mainly for two reasons. First, its existence is testable at the LHC. The direct-detection searches for scalar leptoquarks 
have been putting exclusion bounds on $S_1$ with different
decay hypotheses~\cite{Sirunyan:2018nkj,Sirunyan:2018kzh}.
Second, as mentioned earlier, such a leptoquark can offer
an explanation of some persistent flavour anomalies observed in several experiments. For example, if we consider the anomalies observed in the $B$-meson 
semileptonic decays via charged currents (collectively these show the most significant departure from the SM expectations), 
$S_1$ can provide an explanation if it couples with $\tau$ and neutrino(s) and $b$ and $c$ quarks. 
The direct LHC bounds on such a leptoquark are not very severe but as it has been pointed out in 
Ref.~\cite{Mandal:2018kau}, the present LHC data in the $pp\to\tau\tau / \tau\nu$ channels have actually put constraints on the $S_1$ parameter space relevant 
for explaining the observed $R_{D^{(*)}}$ anomalies. 
Here, using flavour data and LHC constraints, we obtain the allowed parameter space in our model. 
We also point out some possible new search channels at the LHC.
On the flavour side, our primary focus  is on the charged-current 
anomalies observed in the semileptonic $B$ decays in the $R_{D^{(*)}}$ observables.
Hence, as in Ref.~\cite{Mandal:2018kau}, we focus on the interaction terms of $S_1$ that could play a role to address the 
$R_{D^{(*)}}$ anomalies for simplicity.

\subsection{The ${\mathbf S_1}$ model}
\noindent
The single TeV scale $S_1$ leptoquark that originates from the GUT model discussed in Sec.~\ref{sec:model}
transforms under the SM gauge group as $\lt(3,1,-1/3\rt)$.  The low-energy interactions of $S_1$ with the SM 
fields are shown in a compact manner in Eq.~\eqref{eq:lagrangianLQ}.
Below, we display the relevant interaction terms required for our phenomenological analysis, 
\begin{align}
\label{eq:lagS1}
\mc{L} \supset \lt[\lm_{ij}^L\bar{Q}^c_i\lt(i\tau_2\rt)L_j + \lm_{ij}^R\bar{u}^c_i\ell_{Rj}\rt] S_1^{\dag} + \mathrm{H.c.}\ ,
\end{align}
where $Q_i$ and $L_i$ denote the $i$th-generation quark and lepton doublets, respectively and $\lm_{ij}^H$ 
represents the coupling of $S_1$ with a charge-conjugate quark of $i$th generation and a lepton of $j$-th 
generation with chirality $H$. Without any loss of generality, we assume all $\lm$'s are real in our collider analysis
since the LHC data that we consider are insensitive to their complex nature. Also, we only consider mixing among quarks
[Cabibbo-Kobayashi-Maskawa (CKM) mixing] 
and ignore neutrino mixing [Pontecorvo-Maki-Nakagawa-Sakata (PMNS) mixing] completely as all neutrino flavours contribute to the missing energy and hence are not not
distinguishable at the LHC. 
The couplings of $S_1$ to the first-generation 
SM fermions are heavily constrained~\cite{Cai:2017wry}. Hence, we assume $\lm_{1i},\lm_{i1}=0$ in our analysis.
\footnote{However, these couplings can be generated through the CKM mixing. 
We refer the interested readers to Ref.~\cite{Cai:2017wry} for various important flavour-constrains in this regard.} 

\begin{figure*}[t]
\captionsetup[subfigure]{labelformat=empty}
\begin{center}
\subfloat[\quad\quad\quad(a)]{\includegraphics[height=3cm,width=4cm]{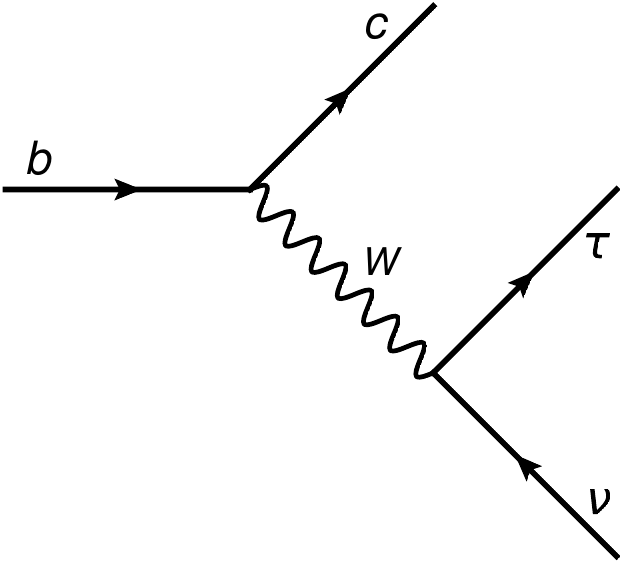}}
\hspace{0.7cm}
\subfloat[\quad\quad\quad(b)]{\includegraphics[height=3cm,width=4cm]{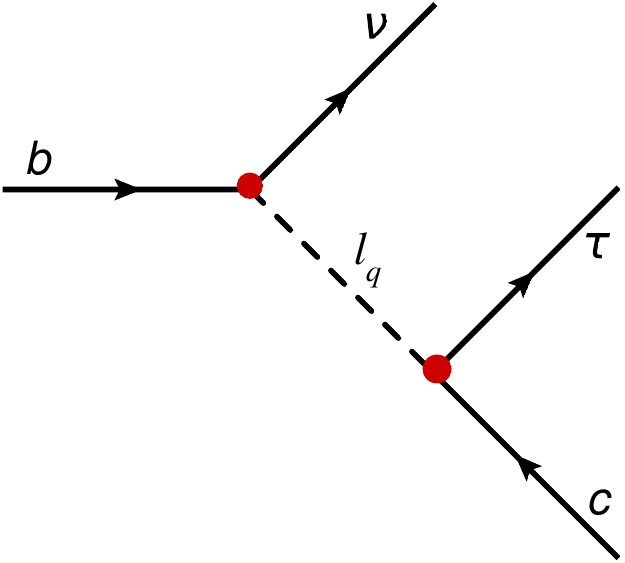}}
\hspace{0.7cm}
\subfloat[\quad\quad\quad(c)]{\includegraphics[height=3cm,width=4cm]{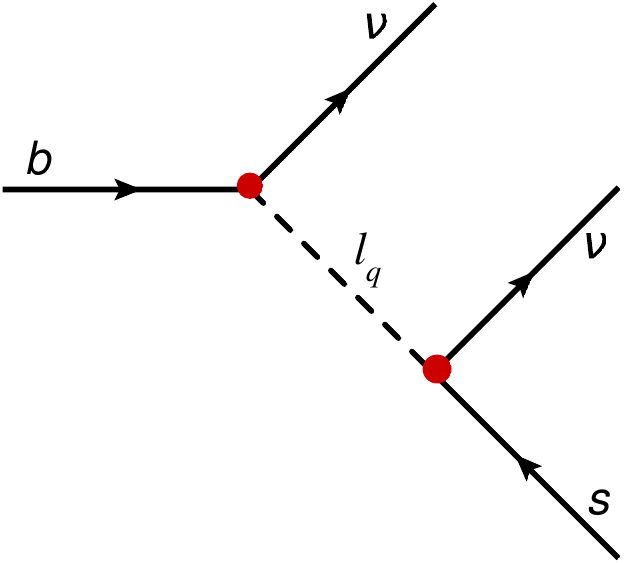}\label{fig:b2svv}}
\caption{Leading-order parton-level Feynman diagrams responsible for the $B\to D^{(*)}\tau\nu$ decay (a) in the 
SM, and (b) in the $S_1$ model. The parton-level process in the presence of $S_1$ that would contribute to $B\to K\nu\nu$ is shown in diagram (c).}
\label{fig:FDRDRDst}
\end{center}
\end{figure*}
The parton level Feynman diagrams for the $b\to c\tau\nu$ decay (responsible for the $B\to D^{(*)}\tau\nu$ decay) are shown in Fig.~\ref{fig:FDRDRDst}.
In order to have a nonzero contribution in the $R_{D^{(*)}}$ observables from $S_1$, we need
$b\nu S_1$ and $c\tau S_1$ couplings to be nonzero simultaneously. Minimally, 
one can  start with just a single free coupling -- either $\lm_{23}^L$ or $\lm_{33}^L$. The coupling $\lm_{23}^L$ 
($\lm_{33}^L$) directly generates $c\tau S_1$ ($b\nu S_1$) interaction and the other one, i.e., 
the $b\nu S_1$ ($c\tau S_1$) can be
generated through the CKM mixing among quarks. These two minimal scenarios were discussed
in detail in Ref.~\cite{Mandal:2018kau}. 
For these two cases, the Lagrangian in Eq.~\eqref{eq:lagS1} can be written explicitly as,
\begin{align}
C_1:~~\mc{L} &\supset \lm_{23}^L\left[\bar c^c\tau_L - \lt(V_{cb}\bar b^c + V_{cs}\bar s^c + V_{cd}\bar d^c\rt)\nu\right]S_1^{\dag}    \nn\\
 &+\ \mathrm{H.c.}\quad(\lm_{33}^L=0),\\
C_2:~~\mc{L} &\supset \lm_{33}^L\left[\lt(V^*_{ub}\bar{u}^c+V^*_{cb}\bar{c}^c+V^*_{tb}\bar{t}^c\rt)\tau_L -\bar{b}^c\nu\;\right] S_1^{\dag}\nn\\ &+\ \mathrm{H.c.}\quad (\lm_{23}^L=0).
\end{align} 
In Ref.~\cite{Mandal:2018kau}, it was shown that for $C_1$, the $R_{D^{(*)}}$-favoured parameter 
space is already ruled out by the latest LHC data. On the other hand, $C_2$ is not seriously constrained by the LHC data since this scenario is 
insensitive to the coupling $\lm_{33}^L$. 
Only the pair-production searches, which are largely insensitive to $\lm_{33}^L$, in the $tt\tau\tau$ and $bb\nu\nu$ modes 
exclude $M_{S_1}$ up to 900 GeV~\cite{Sirunyan:2018nkj} and 1100 GeV~\cite{Sirunyan:2018kzh},
respectively for a 100\% BR in each decay mode. However, Ref.~\cite{Bansal:2018nwp}
showed that the $R_{D^{(*)}}$-favoured parameter 
space in $C_2$ is also ruled out by the electroweak precision data on the $Z\to\tau\tau$ decay.

The above two minimal cases, $C_1$ and $C_2$, are the two extremes. One can, however, consider a next-to-minimal situation, where both $\lm_{23}^L$
and $\lm_{33}^L$ are nonzero to explain $R_{D^{(*)}}$ anomalies being within the LHC 
bounds~\cite{Mandal:2018kau}. However, $B\to K^{(*)}\nu\nu$ decay results severely constrain such a scenario due to tree-level
leptoquark contribution [see Fig.~\ref{fig:b2svv}]. References~\cite{Cai:2017wry,Angelescu:2018tyl} indicated that a large $\lm_{23}^R$ might help explain various flavour anomalies simultaneously
while being consistent with other relevant experimental results. 
\bigskip

In this paper, we allow $\lm_{23}^L$, $\lm_{33}^L$ and $\lm_{23}^R$ to be nonzero and perform a parameter scan for a single $S_1$ 
solution of the $R_{D^{(*)}}$ anomalies. We locate the $R_{D^{(*)}}$ favoured parameter space 
that satisfies the limits from $B\to K^{(*)}\nu\nu$ and $Z\to\tau\tau$ decays and is still allowed by the 
latest LHC data. An $S_1$ can also provide new final states at the LHC like
$\tau\tau+jets$ and $\tau+\slashed{E}_T+jets$ in which leptoquarks have not been searched for before.

\begin{figure}[t]
\begin{center}
\includegraphics[width=7.5cm]{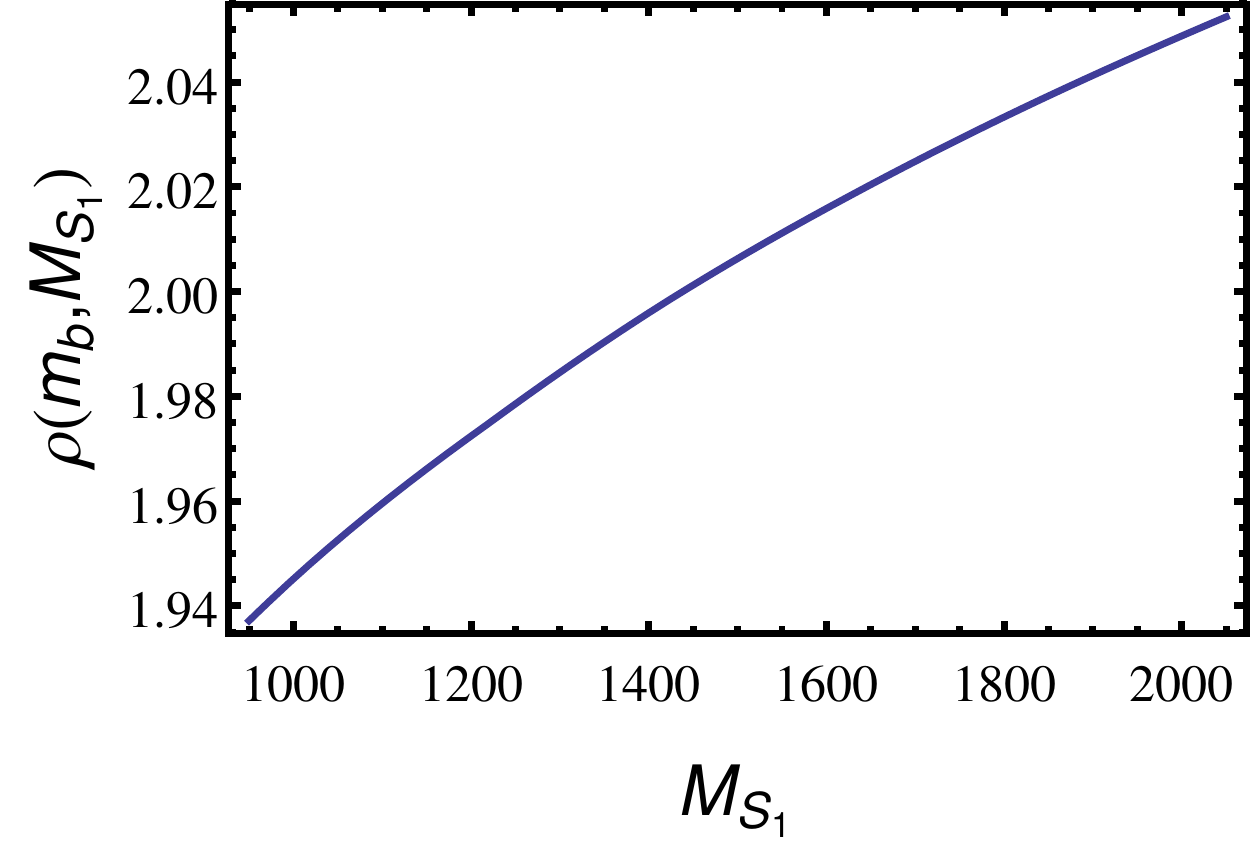}
\caption{The ratio defined in Eq.~\eqref{eq:running}, $ \rho\lt(m_b,M_{S_1}\rt)=\frac{\mc{C}_{S_L}}{\mc{C}_{T_L}}\lt(\mu={m_b}\rt)/ \frac{\mc{C}_{S_L}}{\mc{C}_{T_L}}\lt(\mu=M_{S_1}\rt)$, obtained from Ref.~\cite{Cai:2017wry}.}
\label{fig:CSCT}
\end{center}
\end{figure}

\begin{table*}[t!]
\caption{Summary of the $R_{D^{(*)}}$ related inputs for our parameter scan.}
\begin{center}
\begin{tabular}{c||c|c|c|c}
\hline
$\vphantom{\Big|}$
Observable & Experimental Average & SM Expectation & Ratio & Value \\
\hline\hline
$R_{D}$ 				& $0.340\pm 0.027\pm 0.013$ \cite{Amhis:2016xyh}						& $0.299\pm 0.003$  \cite{Bigi:2016mdz}		& $r_{D}$ & $1.137\pm 0.101$\\
$R_{D^{*}}$ 		& $0.295 \pm 0.011\pm 0.008$ \cite{Amhis:2016xyh} 						& $0.258 \pm 0.005$     \cite{Amhis:2016xyh} 	& $r_{D^{*}}$ & $1.144\pm 0.057$\\
$F_L(D^*)$ 		& $0.60\pm 0.08 \pm 0.035$ \cite{Hirose:2016wfn,Hirose:2017dxl}	& $0.46  \pm 0.04$ \cite{Bhattacharya:2018kig} & $f_L(D^*)$ & $1.313\pm 0.198$\\
$P_\tau(D^*)$ 	& $-0.38 \pm 0.51^{+0.21}_{-0.16}$ \cite{Adamczyk:2019wyt} 		& $-0.497\pm0.013$  \cite{Tanaka:2012nw}	& $p_\tau(D^*)$ & $0.766\pm 1.093$ \\
\hline
\end{tabular}
\label{tab:obsval}
\end{center}
\end{table*}

\subsection{${\mathbf R_{D^{(*)}}}$ with ${\mathbf S_1}$}
\noindent
In the SM, the semitauonic $B$ decay is mediated by the left-handed charged currents and the corresponding 
four-Fermi interactions are given by the following effective Lagrangian
\begin{align}
\mc{L}_{\rm SM} = -\frac{4G_F}{\sqrt{2}}V_{cb}\lt[\bar{c}\gm^\mu P_L b\rt]\lt[\bar{\tau}\gm_\mu P_L\nu_\tau\rt]\, .
\end{align}
In the presence of new physics, there are a total five four-Fermi operators that appear in the effective Lagrangian 
for the $B\to D^{(*)}\tau\nu$ decay~\cite{Tanaka:2012nw},
\begin{align}
\mc{L} \supset -\frac{4G_F}{\sqrt{2}}V_{cb}&\lt[  \lt(1+\mc{C}_{V_L}\rt)\mc{O}_{V_L} + \mc{C}_{V_R}\mc{O}_{V_R} 
+ \mc{C}_{S_L}\mc{O}_{S_R} \rt.\nn\\
&\lt.+ \mc{C}_{S_R}\mc{O}_{S_R} + \mc{C}_{T_R}\mc{O}_{T_R}\rt]\ ,
\end{align}
where the $\mc{C}_X$'s are the Wilson coefficients associated with the effective operators:
\begin{itemize}
\item Vector operators: \vspace{-0.35cm}
$$\begin{array}{ccc}
\mc{O}_{V_L} &=& \lt[\bar{c}\gm^\mu P_L b\rt]\lt[\bar{\tau}\gm_\mu P_L\nu\rt],\\
\mc{O}_{V_R} &=& \lt[\bar{c}\gm^\mu P_R b\rt]\lt[\bar{\tau}\gm_\mu P_L\nu\rt].
\end{array}$$

\item \vspace{-0.35cm} Scalar operators: \vspace{-0.35cm}
$$\begin{array}{cccc}
\mc{O}_{S_L} &=& \lt[\bar{c}P_L b\rt]\lt[\bar{\tau}P_L\nu\rt],&\\
\mc{O}_{S_R} &=& \lt[\bar{c}P_R b\rt]\lt[\bar{\tau}P_L\nu\rt].&
\end{array}$$

\item \vspace{-0.35cm} Tensor operator: \vspace{-0.35cm} 
$$\begin{array}{ccc}
\mc{O}_{T_L} &=& \lt[\bar{c}\sg^{\mu\nu} P_L b\rt]\lt[\bar{\tau}\sg_{\mu\nu} P_L\nu\rt].
\end{array}$$
\end{itemize}

\noindent
The operator
$\mc{O}_{V_L}$ is SM-like and the other four operators introduce new Lorentz structures into the Lagrangian.
Note that the operator $\mc{O}_{T_R}$ is identically zero, i.e.,
\be
\mc{O}_{T_R}=\lt[\bar{c}\sg^{\mu\nu} P_R b\rt]\lt[\bar{\tau}\sg_{\mu\nu} P_L\nu\rt]=0.
\ee

The $S_1$ leptoquark that we consider can generate only  $\mc{O}_{V_L,S_L,T_L}$. Hence, the coefficients of the other two operators, namely, $\mc{C}_{V_R}$ and $\mc{C}_{S_R}$ remain zero in our model. In terms of the $S_1$ parameters the Wilson coefficients can be expressed as,
\begin{eqnarray}
\left.\begin{array}{ll}
\mc{C}_{V_L} &=  \displaystyle\frac{1}{2\sqrt{2}G_FV_{cb}}\frac{\lm_{23}^{L*}\lm_{33}^L}{2M_{S_1}^2}\, , \\
\mc{C}_{S_L} &=  \displaystyle-\frac{1}{2\sqrt{2}G_FV_{cb}}\frac{\lm_{33}^L\lm_{23}^R}{2M_{S_1}^2}\, ,  \\
\mc{C}_{T_L} &= \displaystyle-\frac{1}{4}\mc{C}_{S_L} \, .
\end{array}\right\}\label{eq:cvlcslctl}
\end{eqnarray}
These relations are obtained at the mass scale $M_{S_1}$. However, running of the strong coupling constant
down to $m_b\sim 4.2$ GeV changes these coefficients substantially except for $\mc{C}_{V_L}$ which is protected 
by the QCD Ward identity. As a result, the ratio $\mc{C}_{S_L}/\mc{C}_{T_L}$ becomes,
\be
\lt.\frac{\mc{C}_{S_L}}{\mc{C}_{T_L}}\rt|_{m_b} = \rho\lt(m_b,M_{S_1}\rt)\lt.\frac{\mc{C}_{S_L}}{\mc{C}_{T_L}}\rt|_{M_{S_1}} = -4\rho\lt(m_b,M_{S_1}\rt).\label{eq:running}
\ee
The modification factor $\rho$ can be obtained from
Ref.~\cite{Cai:2017wry}, and we display it in Fig.~\ref{fig:CSCT}. 
In terms of the nonzero Wilson coefficients we can express 
the ratios $r_{D^{(*)}} = R_{D^{(*)}}/R_{D^{(*)}}^{\mathrm{SM}}$
as~\cite{Iguro:2018vqb},
\begin{align}
r_D \equiv \frac{R_{D}}{R_{D}^{\mathrm{SM}}}\approx&\ |1+\mc{C}_{V_L}|^2 + 1.02\ |\mc{C}_{S_L}|^2 + 0.9\ |\mc{C}_{T_L}|^2 \nn\\
&+ 1.49\ \textrm{Re}\lt[(1+\mc{C}_{V_L})\mc{C}_{S_L}^{*}\rt]\nn\\
&+ 1.14\ \textrm{Re}\lt[(1+\mc{C}_{V_L})\mc{C}_{T_L}^{*}\rt], \\&\nn\\
r_{D^{*}} \equiv \frac{R_{D^{*}}}{R_{D}^{\mathrm{SM}}} \approx&\ |1+\mc{C}_{V_L}|^2 + 0.04\ |\mc{C}_{S_L}|^2 + 16.07\ |\mc{C}_{T_L}|^2 \nn\\
&- 0.11\ \textrm{Re}\lt[(1+\mc{C}_{V_L})\mc{C}_{S_L}^{*}\rt]\nn\\
&- 5.12\ \textrm{Re}\lt[(1+\mc{C}_{V_L})\mc{C}_{T_L}^{*}\rt] .
\end{align}
With Eq.~\eqref{eq:running} one can simplify the above equations as,
\begin{align}
r_D = &\ |1+\mc{C}_{V_L}|^2 + \left(1.02+\frac{0.9}{16\rho^2}\rt) |\mc{C}_{S_L}|^2  \nn\\
&+ \left(1.49-\frac{1.14}{4\rho}\rt)\textrm{Re}\lt[(1+\mc{C}_{V_L})\mc{C}_{S_L}^{*}\rt], \label{eq:rdreduced}\\&\nn\\
r_{D^{*}} =&\ |1+\mc{C}_{V_L}|^2 + \left(0.04+\frac{16.07}{16\rho^2}\rt) |\mc{C}_{S_L}|^2 \nn\\
&- \left(0.11-\frac{5.12}{4\rho}\rt)\textrm{Re}\lt[(1+\mc{C}_{V_L})\mc{C}_{S_L}^{*}\rt], \label{eq:rdstreduced}
\end{align}
where $\rho=\rho\lt(m_b,M_{S_1}\rt)$.
There are two other observables related to the $R_{D^*}$--the longitudinal $D^*$ polarization $F_L(D^*)$ and the longitudinal $\tau$ polarization asymmetry $P_{\tau}(D^*)$--have recently been measured by the Belle Collaboration~\cite{Adamczyk:2019wyt,Hirose:2016wfn,Hirose:2017dxl}. 
In terms of the nonzero Wilson coefficients in our model, $F_L(D^*)$ and $P_{\tau}(D^*)$ are expressed as~\cite{Iguro:2018vqb},
\begin{align}
f_L(D^*) \equiv& \frac{F_{L}(D^*)}{F_{L}^{\textrm{SM}}(D^*)} \approx\ \frac{1}{r_{D^{*}}}\Big\{|1+\mc{C}_{V_L}|^2 + 0.08\  |\mc{C}_{S_L}|^2 \nn\\&+ 7.02\  |\mc{C}_{T_L}|^2 
- 0.24\ \textrm{Re}\lt[(1+\mc{C}_{V_L})\mc{C}_{S_L}^{*}\rt]\nn\\ 
&- 4.37\ \textrm{Re}\lt[(1+\mc{C}_{V_L})\mc{C}_{T_L}^{*}\rt] \Big\},\\&\nn\\
p_{\tau}(D^*) \equiv& \frac{P_{\tau}(D^*)}{P_{\tau}^{\textrm{SM}}(D^*)} \approx\ \frac{1}{r_{D^{*}}}\Big\{|1+\mc{C}_{V_L}|^2 - 0.07\ |\mc{C}_{S_L}|^2\nn\\
& - 1.86\times |\mc{C}_{T_L}|^2  + 0.22\ \textrm{Re}\lt[(1+\mc{C}_{V_L})\mc{C}_{S_L}^{*}\rt] \nn\\
&- 3.37\ \textrm{Re}\lt[(1+\mc{C}_{V_L})\mc{C}_{T_L}^{*}\rt] \Big\} .
\end{align}
These equations can further be simplified as,
\begin{align}
f_L(D^*) =&\ \frac{1}{r_{D^{*}}}\Big\{|1+\mc{C}_{V_L}|^2 + \left(0.08+\frac{7.02}{16\rho^2}\rt) |\mc{C}_{S_L}|^2 \nn\\
&- \left(0.24-\frac{4.37}{4\rho}\rt)\textrm{Re}\lt[(1+\mc{C}_{V_L})\mc{C}_{S_L}^{*}\rt]\Big\},\label{eq:flreduced}\\
p_{\tau}(D^*) =&\ \frac{1}{r_{D^{*}}}\Big\{|1+\mc{C}_{V_L}|^2 -\left(0.07+\frac{ 1.86}{16\rho^2}\rt)|\mc{C}_{S_L}|^2\nn\\
&+ \left(0.22+\frac{3.37}{4\rho}\rt)\textrm{Re}\lt[(1+\mc{C}_{V_L})\mc{C}_{S_L}^{*}\rt]\Big\} .\label{eq:ptreduced}
\end{align}
These two observables have the power to discriminate between new physics models with different 
Lorentz structures (see e.g. Ref.~\cite{Blanke:2018yud}). In Table~\ref{tab:obsval}, we list the bounds on  the $R_{D^{(*)}}$ related observables that we include in our parameter scan.

\begin{figure}[t]
\captionsetup[subfigure]{labelformat=empty}
\begin{center}
\includegraphics[height=6.5cm,width=7.0cm]{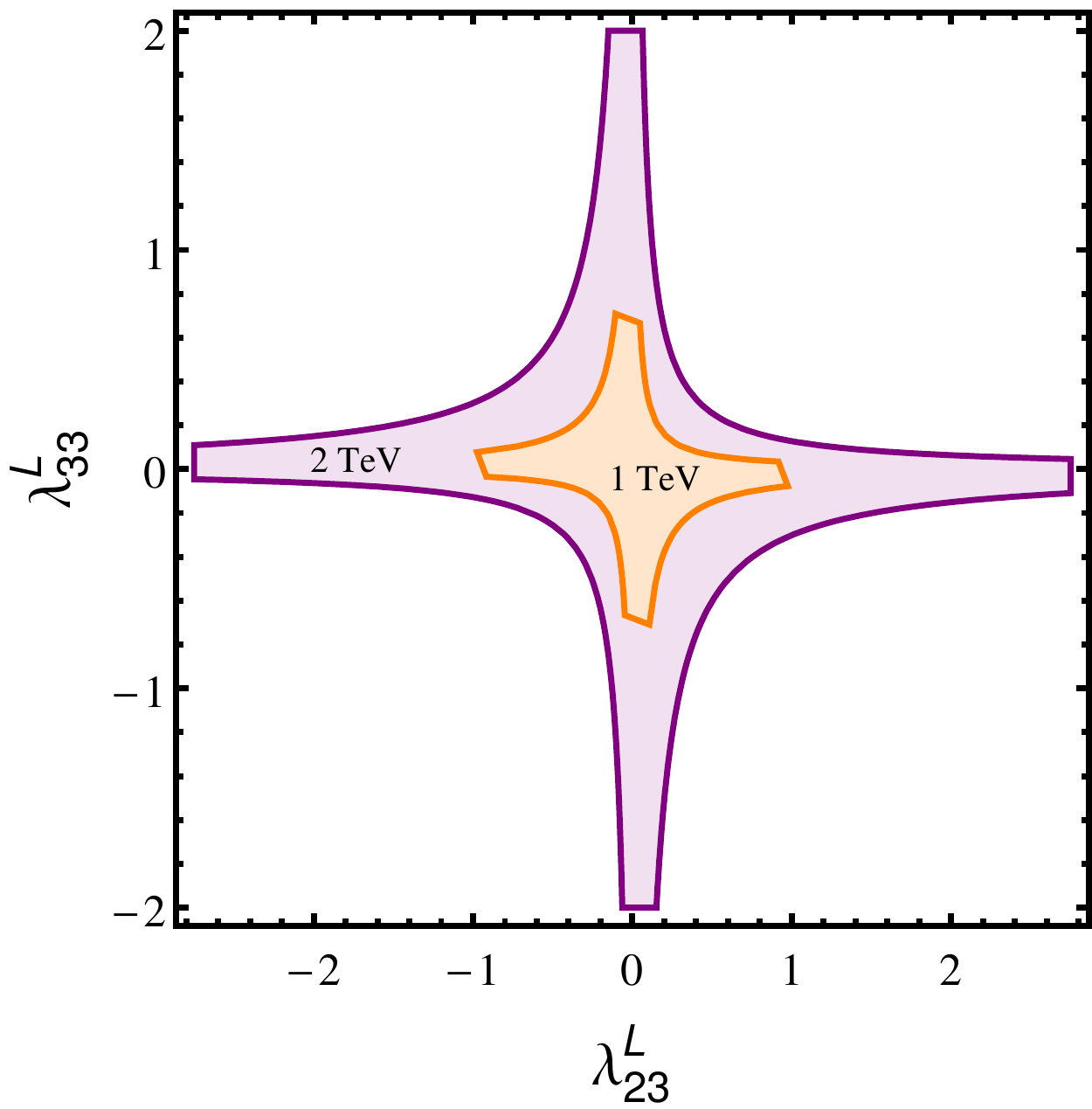}
\caption{\label{fig:RKstvv}The region where $R_{K^{*}}^{\nu\nu}<2.7$ for $M_{S_1}=1$ and $2$ TeV.}
\end{center}
\end{figure}

\subsection{Constraint from ${\mathbf R_{K^{(*)}}^{\nu\nu}}$}
\noindent
The SM flavour-changing neutral-current $b\to s\bar{\nu}\nu$ transition proceeds through a loop and is suppressed by the Glashow-Iliopoulos-Maiani mechanism,
whereas in our model, $S_1$ can mediate this transition at the tree level [see Fig.~\ref{fig:b2svv}].
Therefore, this neutral-current decay can heavily constrain the parameter space of our model. 
We define the following ratio:
\begin{align}
R_{K^{(*)}}^{\nu\nu} = \frac{\Gm(B\to K^{(*)}\nu\nu)}{\Gm(B\to K^{(*)}\nu\nu)_{\mathrm{SM}}}.
\end{align}
The current experimental 90\% confidence limit (C.L.) upper limits on the above quantities are 
$R_{K}^{\nu\nu} < 3.9$ and $R_{K^{*}}^{\nu\nu}<2.7$~\cite{Grygier:2017tzo}.
In terms of
our model parameters, $R_{K^{(*)}}^{\nu\nu}$ is given by the following expression, 
\begin{equation}
R_{K^{(*)}}^{\nu\nu} = 1 - \frac{2a}{3 M_{S_1}^2}\mathrm{Re}\lt(\frac{\lm_{23}^{L*}\lm_{33}^L}{V_{tb}V_{ts}^{*}}\rt)
+ \frac{a^2}{3 M_{S_1}^4}\lt|\frac{\lm_{23}^{L*}\lm_{33}^L}{V_{tb}V_{ts}^{*}}\rt|^2,\label{eq:rkvv}
\end{equation}
where $a=\sqrt{2}\pi^2/\lt(e^2G_F|C_L^{\mathrm{SM}}|\rt)$ with $C_L^{\mathrm{SM}}\approx -6.38$~\cite{Cai:2017wry}.
We use this constraint in our analysis and find that it significantly restricts our parameter space. Note that this
constraint applies on $\lm_{23}^L$ and $\lm_{33}^L$ but not on $\lm_{23}^R$. In Fig.~\ref{fig:RKstvv}, we
show the regions in the $\lm_{23}^L$-$\lm_{33}^L$ plane with $R_{K^{*}}^{\nu\nu}<2.7$ for two different values of $M_{S_1}$.

\subsection{Constraint from $\mathbf Z\to\tau\tau$ decay}
\noindent
Another important constraint comes from the $Z\tau\tau$ coupling measurements. The $Z\to\tau\tau$ decay is
affected by the $S_1$ loops as shown in Ref.~\cite{Bansal:2018nwp}. The contribution
of $S_1$ to the $Z\tau\tau$ coupling shift ($\Dl\kp_{Z\tau\tau}$) comes from a loop with an up-type quark ($q$) and an $S_1$. The shift scales as the square of the $S_1t\tau$ coupling ($\lm^{L/R}_{q3}$) and $m_q^2$. 
Hence, the dominant contribution comes from when $q$ is the top quark implying that the $Z\tau\tau$ coupling
measurements can restrict only $\lm_{33}^L$ but not $\lm_{23}^{L}$ or $\lm_{23}^{R}$.
For instance, we see from Ref.~\cite{Bansal:2018nwp} that $\lm_{33}^L\gtrsim 1.4$ can be excluded for $M_{S_1}\sim 1$ TeV with $2\sg$ confidence. We incorporate this bound into our parameter scan.

\subsection{LHC phenomenology and constraints}
\noindent
We now make a quick survey of the relevant LHC phenomenology of a TeV-range $S_1$ that couples with $\tau$, $\nu$ and $s$ and $c$ quarks. For this discussion we compute all the necessary cross sections using the universal FeynRules output (UFO)
~\cite{Degrande:2011ua} model
files from Ref.~\cite{Mandal:2018kau} in \textsc{MadGraph5}~\cite{Alwall:2014hca}. We use the NNPDF23LO~\cite{Ball:2012cx} 
parton distribution functions (PDFs). Wherever required, we include the next-to-leading-order QCD $K$-factor of $\sim 1.3$ for the pair production in our analysis~\cite{Mandal:2015lca}.

\subsubsection{Decay modes of $S_1$}
\noindent
For nonzero $\lm_{23}^L$, $\lm_{33}^L$ and $\lm_{23}^R$, 
$S_1$ can decay to $c\tau$, $s\nu$, $t\tau$ and $b\nu$ states. CKM mixing among quarks enables decays to $u\tau$ and $d\nu$ but we neglect them in our analysis as the off-diagonal
CKM elements are small. The BRs of $S_1$ to various decay modes vary depending on the coupling strengths. If $\lm_{23}^R\gg \lm_{23}^L,\lm_{33}^L$, the dominant decay mode is $S_1\to c\tau$, whereas
for $\lm_{23}^L\gg \lm_{23}^R,\lm_{33}^L$, $\mathrm{BR}(S_1\to c\tau) \approx \mathrm{BR}(S_1\to s\nu)\approx 50\%$. On the other hand,
when $\lm_{33}^L\gg \lm_{23}^L,\lm_{23}^R$, the dominant decay modes are $S_1\to t\tau$ and $S_1\to b\nu$ with about 50\% BR in each mode. Since partial decay widths depend linearly on $M_{S_1}$, BRs are insensitive to the mass of $S_1$.

\subsubsection{Production of $S_1$}
\noindent
At the LHC, $S_1$ can be produced resonantly in pairs or singly and nonresonantly through
indirect  production ($t$-channel $S_1$ exchange process). 

\vspace{0.25cm}
\noindent
\underline{{\bf \emph{Pair production:}}}
The pair production of $S_1$ is dominated by the strong coupling and, therefore, it is
almost model independent. The mild model dependence enters in the pair production through the $t$-channel
lepton or neutrino exchange processes. However, the amplitudes of those diagrams are proportional
to $\lm^2$ and generally suppressed for small $\lm$ values (for bigger $M_{S_1}$ and large $\lm$ values, this part
could be comparable to the model independent part of the pair production). Pair production is heavily
phase-space suppressed for large $M_{S_1}$ and we find that its contribution is very small in our recast analysis.
Pair production can be categorized into two types depending on the final states: symmetric, where both leptoquarks decay to the same modes, asymmetric, where the two leptoquarks decay via two different modes. These two types give rise to various novel final states.

\begin{widetext}
\noindent
\underline{\emph{Symmetric modes:}}
\begin{align}
S_1S_1 \to c\tau\ubr{-2.5}\ c\tau\ubr{-2.5}\equiv \tau\tau+2j,\quad
t\tau\ubr{-2.5}\ t\tau\ubr{-2.5}\equiv tt+\tau\tau,\quad
s\nu\ubr{-2.5}\ s\nu\ubr{-2.5}\equiv 2j+\slashed{E}_{\rm T},\quad
b\nu\ubr{-2.5}\ b\nu\ubr{-2.5}\equiv 2b+\slashed{E}_{\rm T}\nn.
\end{align}  
\vspace{0.25cm}
\noindent
\underline{\emph{Asymmetric modes:}}
\begin{align}
S_1S_1 &\to c\tau\ubr{-2.5}\ s\nu\ubr{-2.5} \equiv \tau+2j+\slashed{E}_{\rm T},\quad
c\tau\ubr{-2.5}\ b\nu\ubr{-2.5} \equiv \tau+b+j+\slashed{E}_{\rm T},\quad
c\tau\ubr{-2.5}\ t\tau\ubr{-2.5} \equiv \tau\tau+t+j \nn,\\
S_1S_1 &\to s\nu\ubr{-2.5}\ b\nu\ubr{-2.5} \equiv b+j+\slashed{E}_{\rm T},\quad
s\nu\ubr{-2.5}\ t\tau\ubr{-2.5} \equiv t+\tau+j+\slashed{E}_{\rm T},\quad
b\nu\ubr{-2.5}\ t\tau\ubr{-2.5} \equiv t+\tau+b+\slashed{E}_{\rm T}\nn,
\end{align} 
\end{widetext}

\noindent where the curved connection over a pair of particles indicates that the pair is coming from a decay of $S_1$. Searches for leptoquarks in some of the symmetric modes were already done at the 
LHC~\cite{Sirunyan:2018nkj,Sirunyan:2018kzh}. Leptoquark searches in
some of the symmetric and most of the asymmetric modes are yet to be performed at the LHC. 

\vspace{0.25cm}
\noindent
\underline{{\bf \emph{Single production:}}}
The single productions of $S_1$, where $S_1$ is produced in association with a SM particle, are fully model dependent as 
they depend on the leptoquark-quark-lepton couplings. These are important production modes for large couplings and heavier masses 
(since single productions receive less phase-space suppression than the pair production). 
Depending on the final states, single productions can be categorized as follows.

\begin{widetext}
\vspace{0.25cm}\noindent \underline{\emph{Symmetric modes:}} 
\begin{eqnarray}
\mc{S}_1\ \tau\ X&\to\  \tau c\ubr{-2.5}\ \tau\ +\ \tau c\ubr{-2.5}\ \tau\ j&\  (+\ \tau c\ubr{-2.5}\ \tau\ jj\ +\ \cdots)\ \equiv\  \tau\tau+jets\nn\\
\mc{S}_1\ \tau\ X&\to\   \tau t\ubr{-2.5}\ \tau\ +\ \tau t\ubr{-2.5}\ \tau\ j&\ (+\ \tau t\ubr{-2.5}\ \tau\ jj\ +\ \cdots)\ \equiv\  \tau\tau+t+jets\nn\\
\mc{S}_1\ \nu\ X&\to\   \nu s\ubr{-2.5}\ \nu\ +\ \nu s\ubr{-2.5}\ \nu\ j&\ (+\ \nu s\ubr{-2.5}\ \nu\ jj\ +\ \cdots)\ \equiv\ \slashed{E}_{\rm T}+jets\nn\\
\mc{S}_1\ \nu\ X&\to\   \nu b\ubr{-2.5}\ \nu\ +\ \nu b\ubr{-2.5}\ \nu\ j&\ (+\ \nu b\ubr{-2.5}\ \nu\ jj\ +\ \cdots)\ \equiv\  \slashed{E}_{\rm T}+b+jets\nn
\end{eqnarray}
\underline{\emph{Asymmetric modes:}}
\begin{eqnarray}
S_1\ \tau \ X  &\to\   \nu s\ubr{-2.5}\ \tau\ +\ \nu s\ubr{-2.5}\ \tau\ j&\ ( +\ \nu s\ubr{-2.5}\ \tau\ jj\ +\ \cdots)\ \equiv\  \slashed{E}_{\rm T}+\tau+jets\nn\\
S_1\ \nu\ X  &\to\    \tau c\ubr{-2.5}\ \nu\ +\ \tau c\ubr{-2.5}\ \nu\ j&\ ( +\ \tau c\ubr{-2.5}\ \nu\ jj\ +\ \cdots)\ \equiv\  \slashed{E}_{\rm T}+\tau+jets\nn\\
S_1\ \tau \ X &\to\   \nu b\ubr{-2.5}\ \tau\ +\ \nu b\ubr{-2.5}\ \tau\ j&\ ( +\ \nu b\ubr{-2.5}\ \tau\ jj\ +\ \cdots)\ \equiv\  \slashed{E}_{\rm T}+\tau+b+jets\nn\\
S_1\ \nu\ X  &\to \ \tau t\ubr{-2.5}\ \nu\ +\ \tau t\ubr{-2.5}\ \nu\ j&\ (+\ \tau t\ubr{-2.5}\ \nu\ jj\ +\ \cdots)\ \equiv\ \slashed{E}_{\rm T}+\tau+t+jets\nn
\end{eqnarray}
\end{widetext}
Here $j$ stands for an untagged jet and $jets$ means any number ($\geq1$) of untagged jets. These extra jets can be either radiation or hard (genuine three-body single production processes can have sizeable cross sections; see Refs.~\cite{Mandal:2012rx,Mandal:2015vfa,Mandal:2016csb} for how one can systematically compute them). As single production is model dependent, the relative strengths of these modes depend on the relative strengths of the coupling involved in the production as well as the BR of the decay mode involved.

\vspace{0.25cm}
\noindent
\underline{{\bf \emph{Indirect production:}}}
Indirect production is the nonresonant process where a leptoquark is exchanged in the $t$ channel. With leptoquark couplings to $\tau$ and $\nu$, this basically
gives rise to three possible final states: $\tau\tau~(\tau\tau+jets)$, $\tau\nu~(\slashed E_{\rm T}+\tau+jets)$ and $\nu\nu~(\slashed E_{\rm T}+jets)$. The amplitudes of these processes are
proportional to $\lm^2$. So the cross section grows as $\lm^4$. Hence, for an order-one $\lm$, indirect production has a larger cross 
section than other production processes for large $M_{S_1}$ (see Fig.~2 of Ref.~\cite{Mandal:2018kau}). However, the indirect production substantially
interferes with the SM background process $pp\to V^{(*)}\to\ell\ell$ ($\ell =\tau/\nu$). Though the interference is $\mc O(\lm^2)$, 
its contribution can be significant for a TeV-scale $S_1$  because of the large SM contribution (larger than both the direct production modes and the $\lm^4$
 indirect contribution, assuming $\lm\gtrsim1$).
In general, the interference could be either constructive or destructive depending on the nature of the leptoquark species and its mass~\cite{Bansal:2018eha}.
For $S_1$, we find that the interference is destructive in nature~\cite{Mandal:2018kau}. Hence, for a TeV-scale $S_1$ if $\lm$ is large, this destructive interference 
becomes its dominant signature in the leptonic final states.

\subsubsection{Constraints from the LHC}
\noindent
The mass exclusion limits from the pair-production searches for $S_1$ at the LHC are as follows. Assuming a $100\%$ BR in the $S\to t \tau$ mode, 
a recent search at the CMS detector has excluded masses below $900$ GeV~\cite{Sirunyan:2018nkj}. Similarly, for a leptoquark that decays exclusively to 
$b\nu$ or $s\nu~(\equiv j\nu$) final states, the exclusion limits are at  $1100$ and $980$ GeV~\cite{Sirunyan:2018kzh}, respectively. 
However, going beyond simple mass exclusions, we make use of the analysis done in Ref.~\cite{Mandal:2018kau} for the LHC constraints. It contains 
the independent LHC limits on the three couplings shown in Eq.~\eqref{eq:lagS1} as functions of $M_{S_1}$ as well as a summery 
of the direct-detection exclusion limits. 

\begin{figure}[t]
\centering
\includegraphics[height=5.5cm,width=7.5cm]{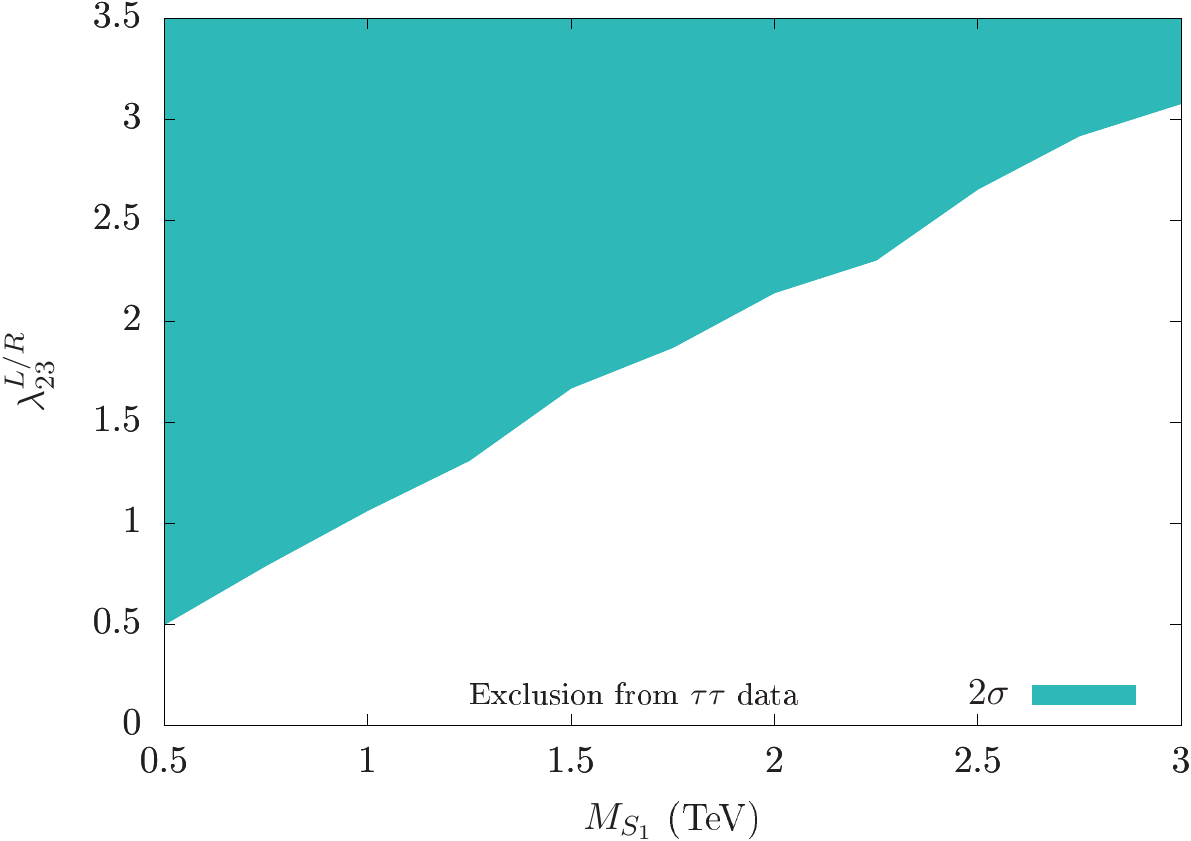}
\caption{Two-sigma exclusion limits on $\lm^{L/R}_{23}=\sqrt{\lt(\lm_{23}^L\rt)^2+\lt(\lm_{23}^R\rt)^2}$  as a function of $M_{S_1}$ as obtained in Ref.~\cite{Mandal:2018kau} from the ATLAS $pp\to \tau\tau$ data~\cite{Aaboud:2017sjh}. The coloured region is excluded.}
\label{fig:l23Lexclu}
\end{figure}

Apart from the processes with $\slashed  E_{\rm T}+jets$ final states, all other production processes can have either 
$\tau\tau+jets$ final states or $\slashed E_{\rm T}+\tau+jets$ final states. Hence, the latest $pp\to Z'\to\tau\tau$ and $pp\to W'\to \tau\nu$ searches at the ATLAS detector~\cite{Aaboud:2017sjh,Aaboud:2018vgh} were used to derive the constraints in Ref.~\cite{Mandal:2018kau}. 
There we notice that the limits on $\lm^L_{23}$ from the $\tau\nu$ data are weaker that the ones obtained from the $\tau\tau$ data. The $\tau\tau$ data also constrain  $\lm^{R}_{23}$.
From the earlier discussion, it is clear that the interference contribution plays the dominant role in determining these limits. However, its destructive nature
means that in the signal region one would expect less events than the SM-only predictions. Hence, the limits were obtained 
assuming that either $\lm^L_{23}$ or $\lm^R_{23}$ is nonzero at a time or by performing a $\chi^2$ 
test of the transverse mass ($m_{\rm T}$) distributions of the data.
As, for heavy $S_1$, the limits on $\lm^L_{23}$ and $\lm^R_{23}$ are dominantly determined by the interference 
of the indirect production, they are very similar. We can translate these limits from the $\tau\tau$ data on any combination of $\lm^L_{23}$ and $\lm^R_{23}$ in a simple manner assuming 
$\lm^{L/R}_{23}=\sqrt{\lt(\lm_{23}^L\rt)^2+\lt(\lm_{23}^R\rt)^2}$. In Fig.~\ref{fig:l23Lexclu} we display the limits on 
$\lm^{L/R}_{23}$ as a function of $M_{S_1}$.\footnote{Actually, for $M_{S_1}$ between $1$ and $2$ TeV, the limits on 
$\lm^L_{23}$ are slightly stronger
than  those on $\lm^R_{23}$ because in the SM the $Z$ boson couples differently to left- and right-handed $\tau$'s. However, we ignore this minor difference 
and take the stronger limits on $\lm^L_{23}$ as the limits on $\lm^{L/R}_{23}$ to remain conservative.}

The LHC data is insensitive to $\lm^L_{33}$ as it was shown in Ref.~\cite{Mandal:2018kau}. This
can be understood from the following argument. First, the pair production is insensitive to this coupling as we have already mentioned. Second, the single-production process $pp\to t\tau\nu$ via an $S_1$ has too small a cross section ($\sim 2$ fb for $\lm^L_{33}\sim 1$ and $M_{S_1}=1$ TeV) to make any difference at the present luminosity. Finally, there is no interference contribution
in the $\tau\tau$ and $\tau+\slashed{E}_{\rm T}$ channels as there is no $t$ quark in the initial state.
Hence, $\lm^L_{33}$ remains unbounded from these searches.

\begin{figure*}[t]
\captionsetup[subfigure]{labelformat=empty}
\begin{tabular}{rrr}
\subfloat[\quad\quad\quad(a)]{\includegraphics[height=5.4cm,width=5.4cm]{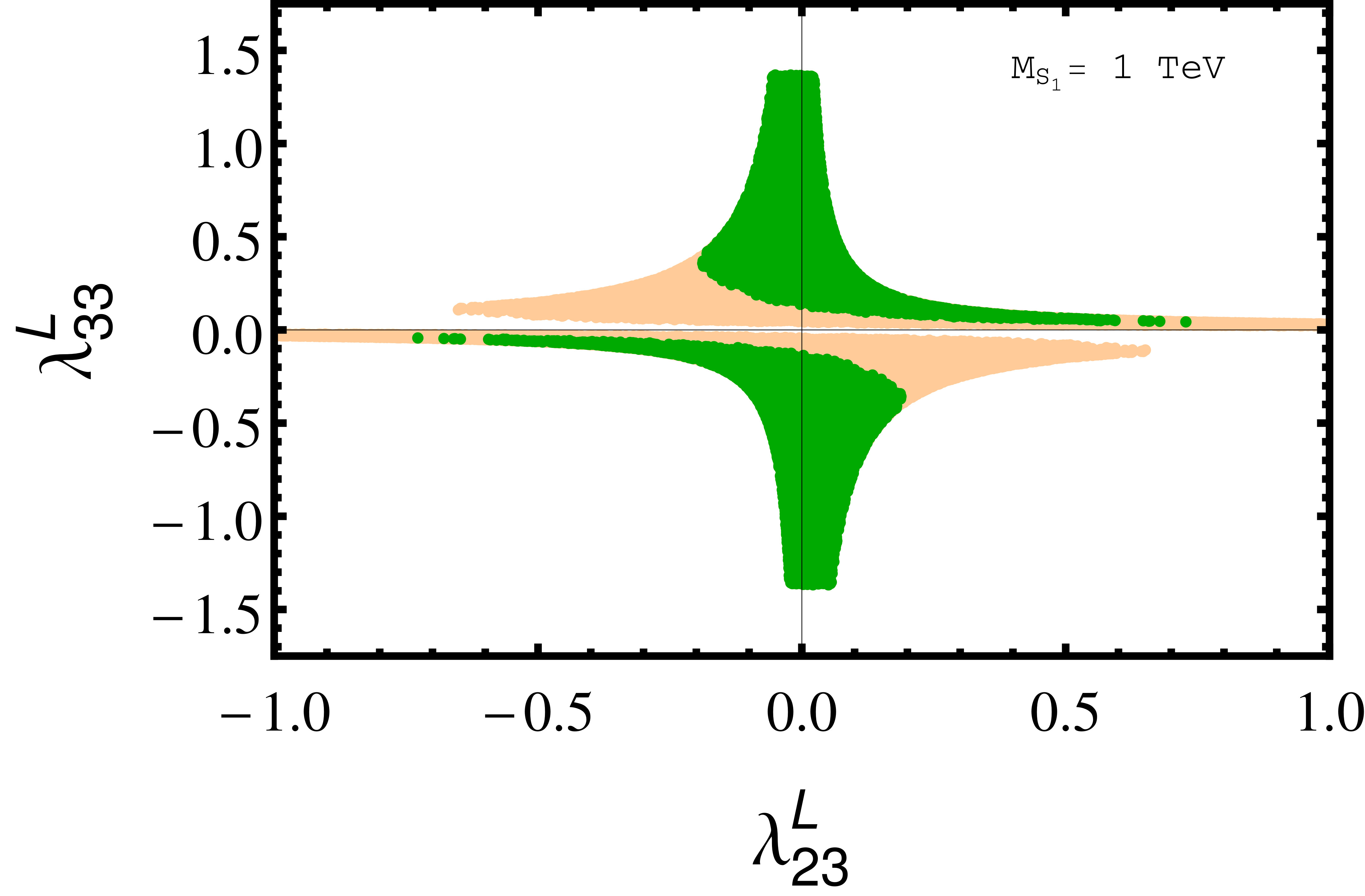}\label{fig:ps23L33La}}&
\subfloat[\quad\quad\quad(b)]{\includegraphics[height=5.4cm,width=5.4cm]{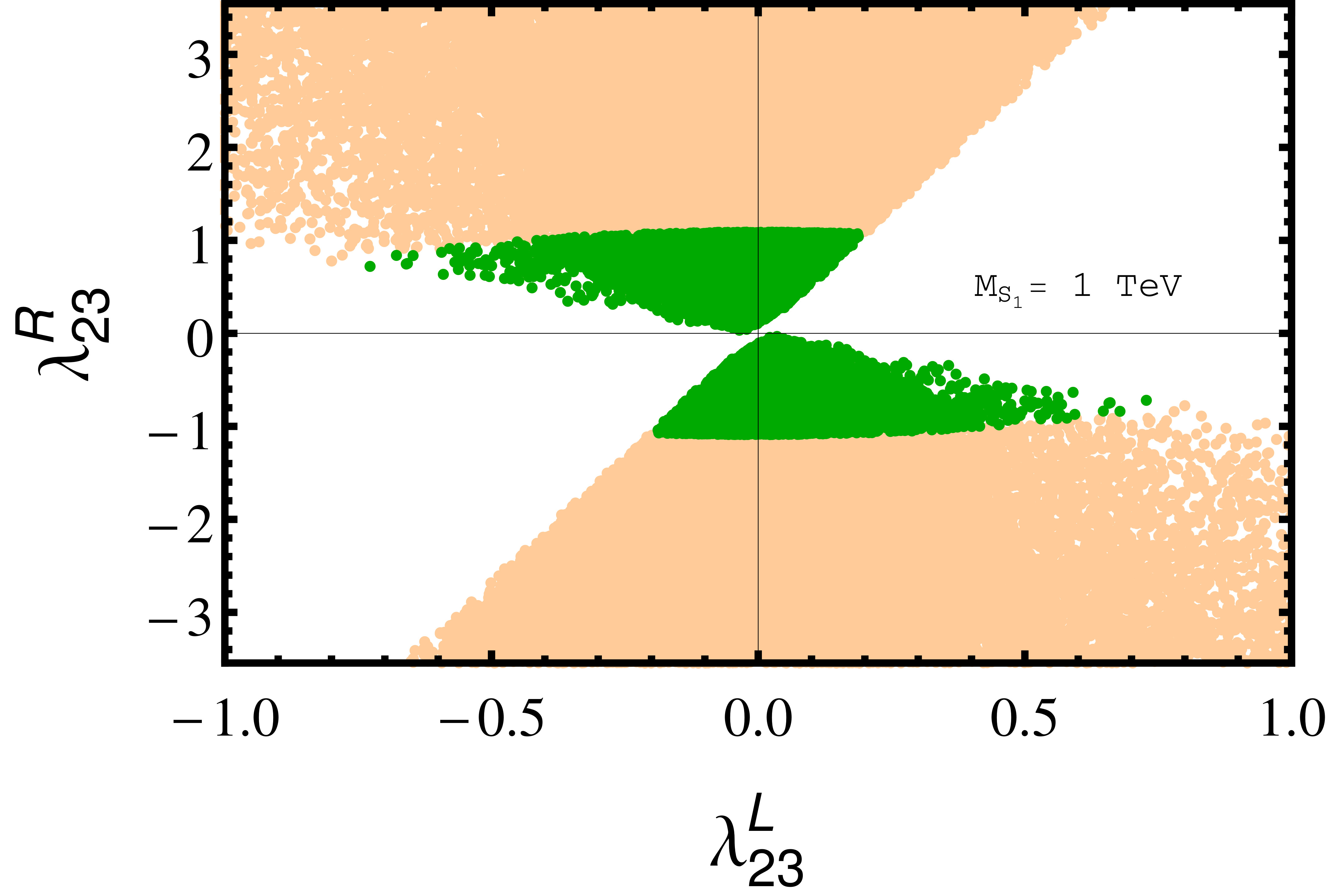}\label{fig:ps23L23Ra}}&
\subfloat[\quad\quad\quad(c)]{\includegraphics[height=5.4cm,width=5.4cm]{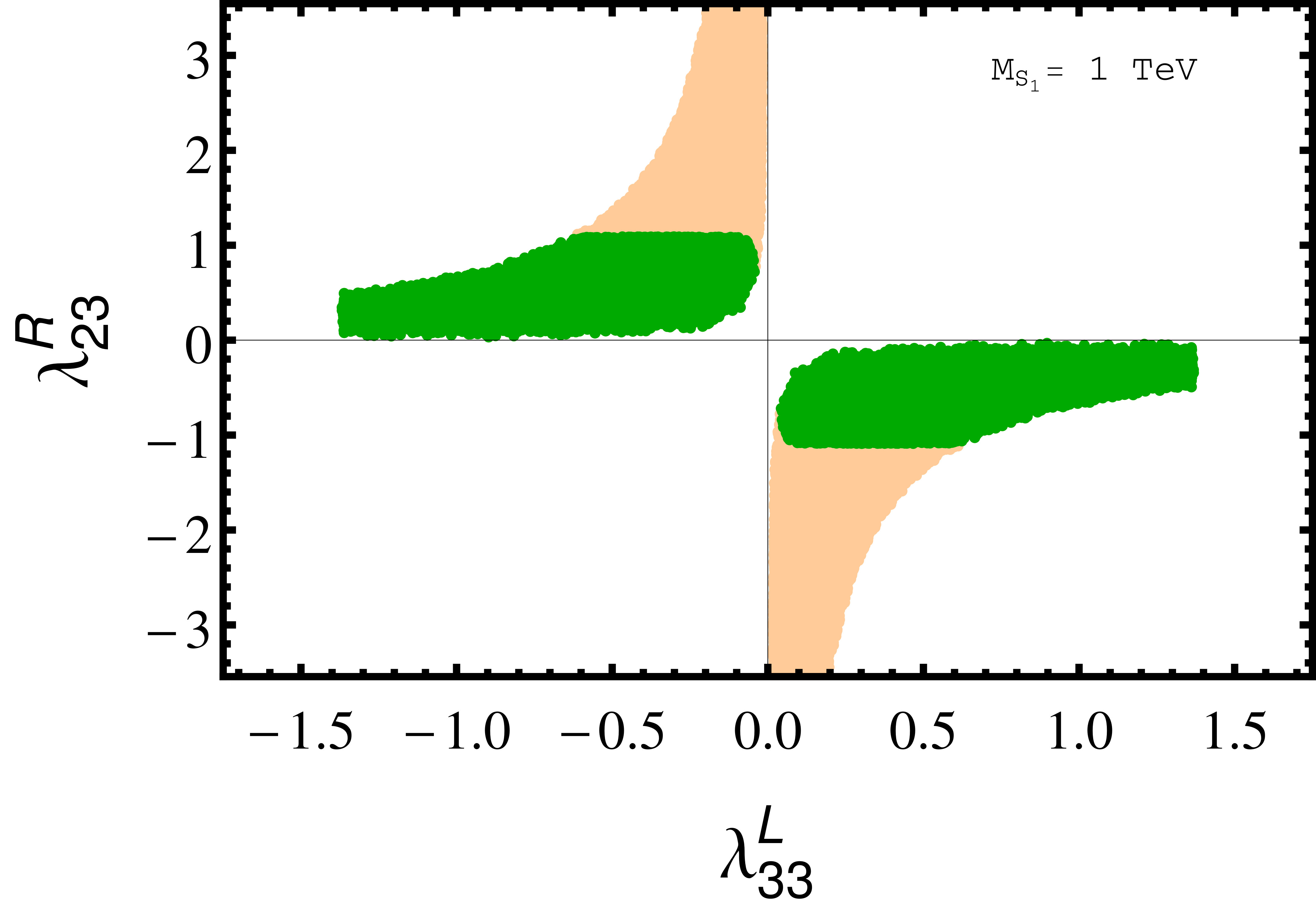}\label{fig:ps33L23Ra}}\\
\subfloat[\quad\quad\quad(d)]{\includegraphics[height=5.4cm,width=5.4cm]{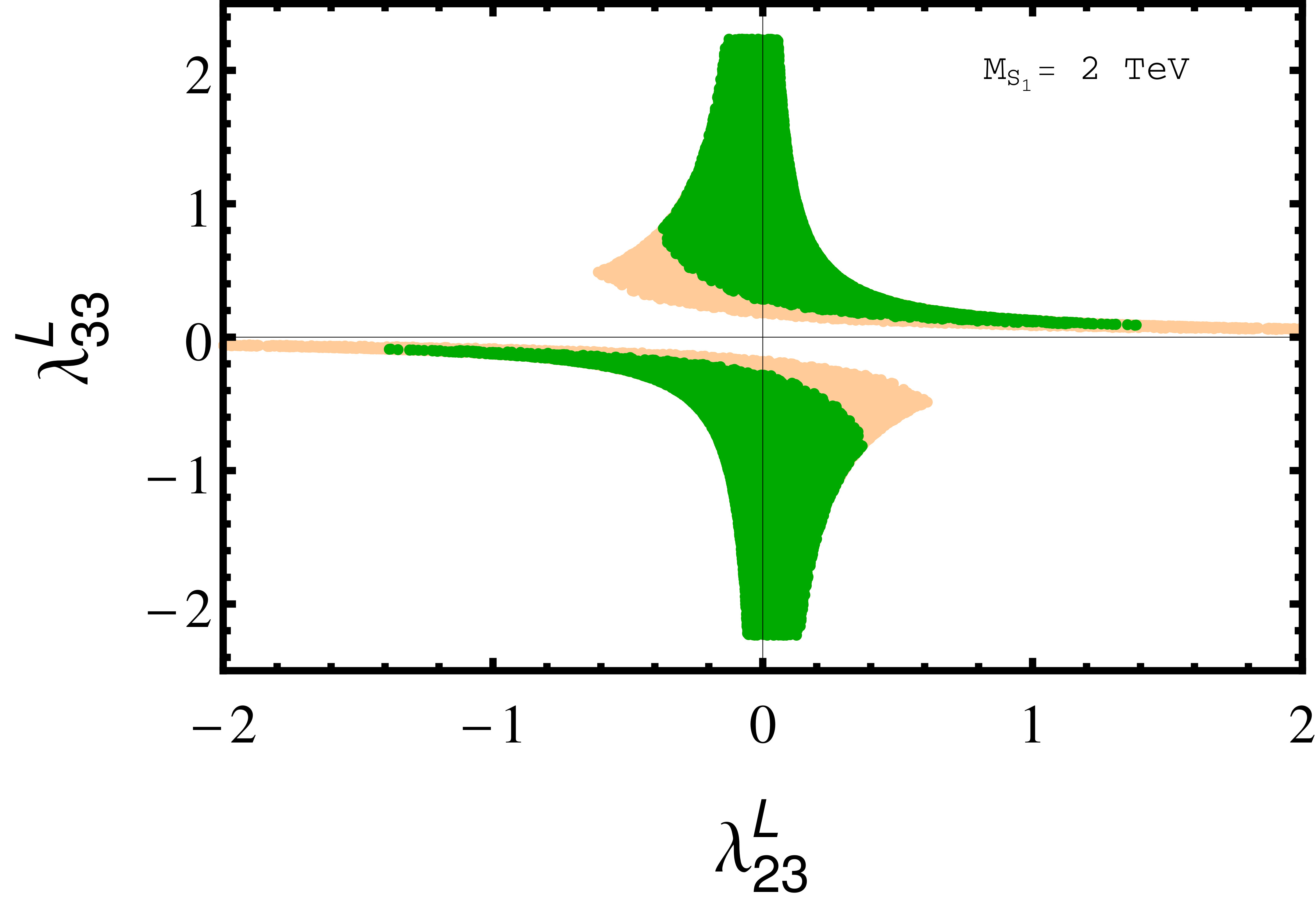}\label{fig:ps23L33Lc}}&
\subfloat[\quad\quad\quad(e)]{\includegraphics[height=5.4cm,width=5.4cm]{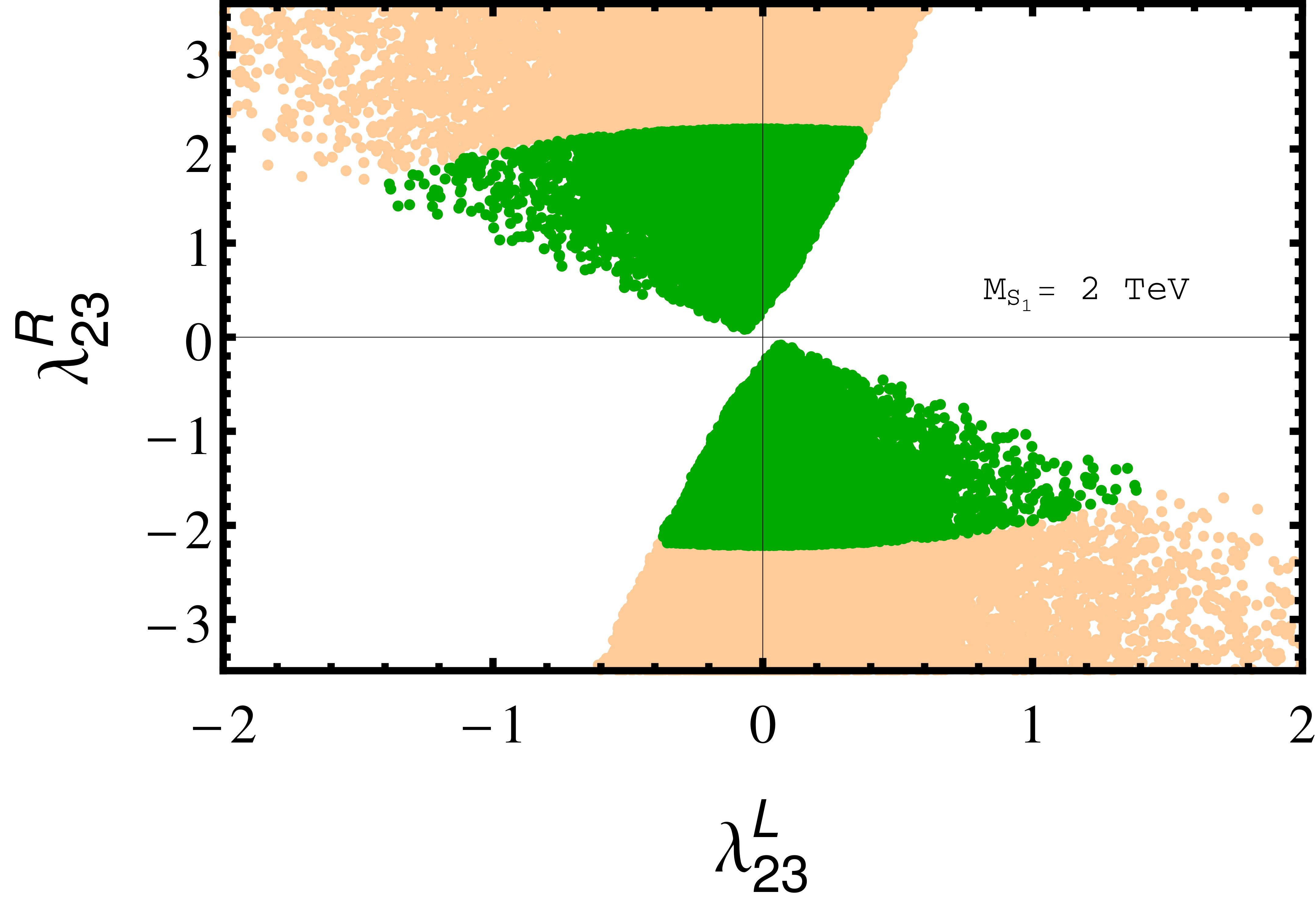}\label{fig:ps23L23Rc}}&
\subfloat[\quad\quad\quad(f)]{\includegraphics[height=5.4cm,width=5.4cm]{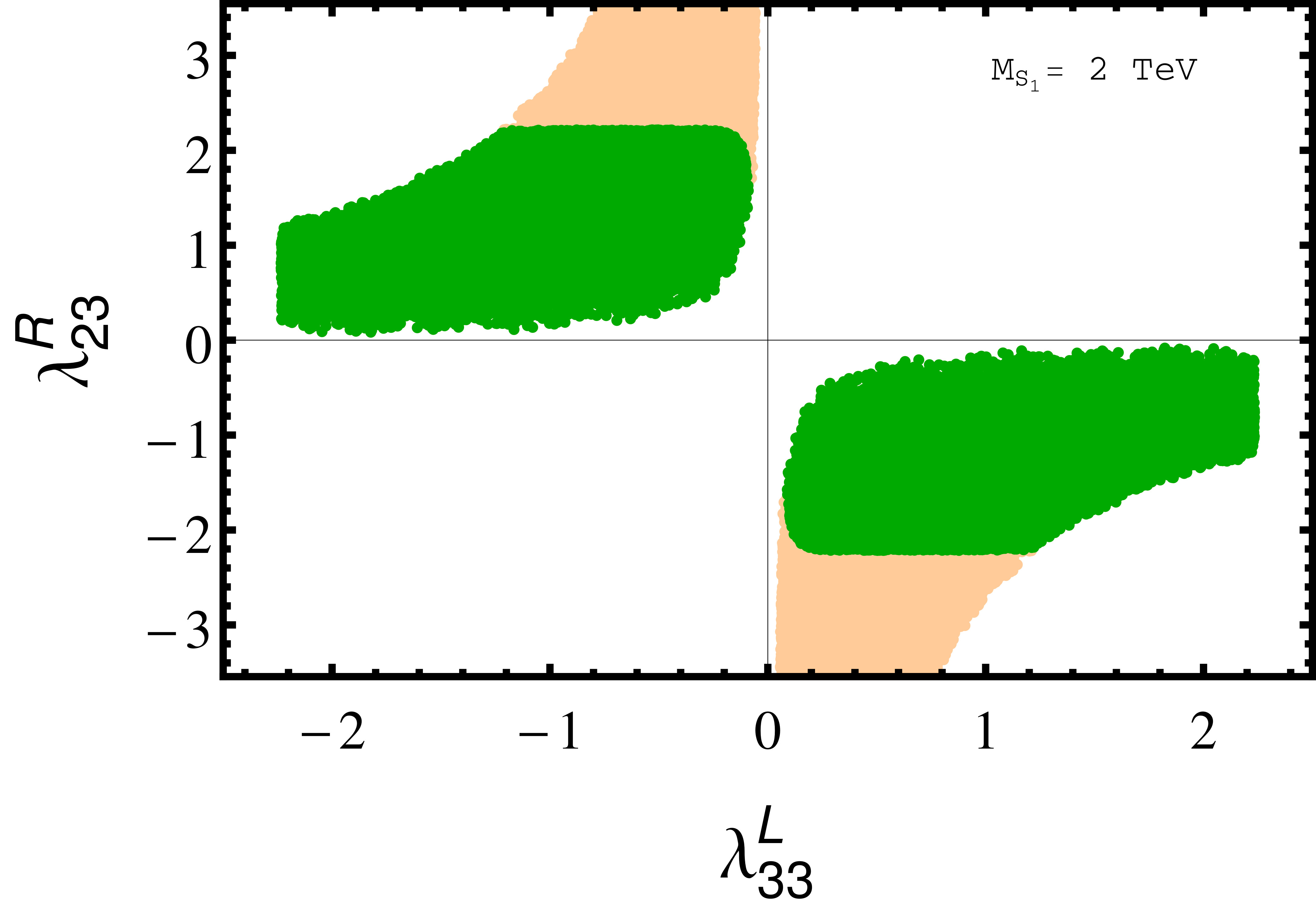}\label{fig:ps33L23Rc}}
\end{tabular}
\caption{Two-dimensional projections of the regions in the $S_1$ parameter space allowed by 
 the bounds on the $R_{D^{(*)}}$, $F_L(D^*)$, $P_\tau(D^*)$ and $R^{\nu\nu}_{K^{(*)}}$ observables and the $Z\to\tau\tau$ decay, i.e., the {\em FEW} regions (orange), see Sec.~\ref{sec:param_scan}, and the LHC constraints in addition to all these constraints, i.e., the {\it FEWL} regions (green) for $M_{S_1}=1$
TeV (upper panel) and $M_{S_1}=2$ TeV (lower panel). We assume all the couplings are real.}
\label{fig:parascan}
\end{figure*}

\subsection{Parameter scan}\label{sec:param_scan}
\noindent
To find the $R_{D^{(*)}}$-favoured regions in the $S_1$ parameter space that is not in conflict with the limits on
$F_L(D^*)$, $P_\tau(D^*)$,
$R^{\nu\nu}_{K^{(*)}}$, $Z\to\tau\tau$ decay 
and the bounds from the LHC, we consider two benchmark leptoquark masses, $M_{S_1}=1$ and $2$ TeV. 
We allow all the three free couplings, $\lm_{23}^L$, $\lm_{23}^R$ and $\lm_{33}^L$ to vary. For every benchmark mass, we perform a random scan over the three couplings in the perturbative range $-\sqrt{4\pi}$ to 
$\sqrt{4\pi}$ (i.e., $|\lm|^2/4\pi \leq 1$). We do not consider
  complex values for the couplings.
In Fig.~\ref{fig:parascan}, we show the outcome of our scan with different two-dimensional projections. 
In every plot we show two couplings and allow the third coupling to vary. 
In each of these plots we show the following.
\begin{itemize}
\item \underline{The {\it {\textbf F}lavour ${\boldsymbol \cup}$ \textbf{EW}} (FEW) regions:} The orange dots mark the regions favoured by the 
$R_{D^{(*)}}$ observables within $95$\% C.L. while satisfying the available bounds on the 
$F_L(D^*)$, $P_\tau(D^*)$ and $R^{\nu\nu}_{K^{(*)}}$ observables (flavour bounds). In addition, these points also satisfy the bound on 
$\lm^{L}_{33}$ coming from the $Z\to\tau\tau$ decay within $95$\% C.L.~\cite{Bansal:2018nwp} (electroweak bound).

\item \underline{The {\it \textbf{F}lavour ${\boldsymbol \cup}$ \textbf{EW} ${\boldsymbol \cup}$ {\textbf L}HC} (FEWL) regions:} As we take into account the limits on $\lm^{L/R}_{23}$ from the ATLAS $pp\to\tau\tau$ data from the $13$ TeV LHC~\cite{Mandal:2018kau} along with the previous constraints we obtain the regions marked by the green points. These are the points that survive all the limits considered in this paper.
\end{itemize}
From the plots we see that substantial portions of parameter regions survive after all the constraints. This implies that the $S_1$ model can successfully explain the $R_{D^{(*)}}$ anomalies.
If one looks only at the $R_{D^{(*)}}$ anomalies, in principle, one can just set $\lm^L_{23}$ and/or $\lm^L_{33}$ to be large. But  coupling values that make $C_{V_L}$ [see Eq.~\eqref{eq:cvlcslctl}] big come into conflict with the $R_{K^{(*)}}^{\n\n}$ bound [see Eq.~\eqref{eq:rkvv}]. This is why we do not see any point where both $\lm^L_{23}$ and $\lm^L_{33}$ are large in the first column of Fig.~\ref{fig:parascan}. In addition, the LHC puts bounds on $\lm^{L/R}_{23}$~\cite{Mandal:2018kau} whereas the $Z\to\tau\tau$ data puts a complimentary bound on $\lm^L_{33}$~\cite{Bansal:2018nwp}.
The restriction on $\lm^{L}_{33}$ from the $Z\to\tau\tau$ data can be seen in Figs.~\ref{fig:ps23L33La} and \ref{fig:ps33L23Ra} for $M_{S_1}=1$ TeV and Figs.~\ref{fig:ps23L33Lc} and \ref{fig:ps33L23Rc} for $M_{S_1}=2$ TeV, whereas from the middle column [i.e., 
Figs.~\ref{fig:ps23L23Ra} and \ref{fig:ps23L23Rc}]
it is clear that the LHC prevents both $\lm^{L}_{23}$ and $\lm^{R}_{23}$ from taking large values simultaneously. From the $\lm^R_{23}$ vs $\lm^L_{33}$ plots [Figs.~\ref{fig:ps33L23Ra} and \ref{fig:ps33L23Rc}] we see that these two couplings take opposite signs mainly because both $R_{D^*}$ and $F_L\lt(D^*\rt)$ prefer a positive $C_{S_L}$ [see Eqs.~\eqref{eq:rdstreduced} and \eqref{eq:flreduced}].

\begin{figure*}[t!]
\captionsetup[subfigure]{labelformat=empty}
\centering
\begin{tabular}{rrr}
\subfloat[\quad\quad\quad(a)]{\includegraphics[height=4.5cm,width=5.4cm]{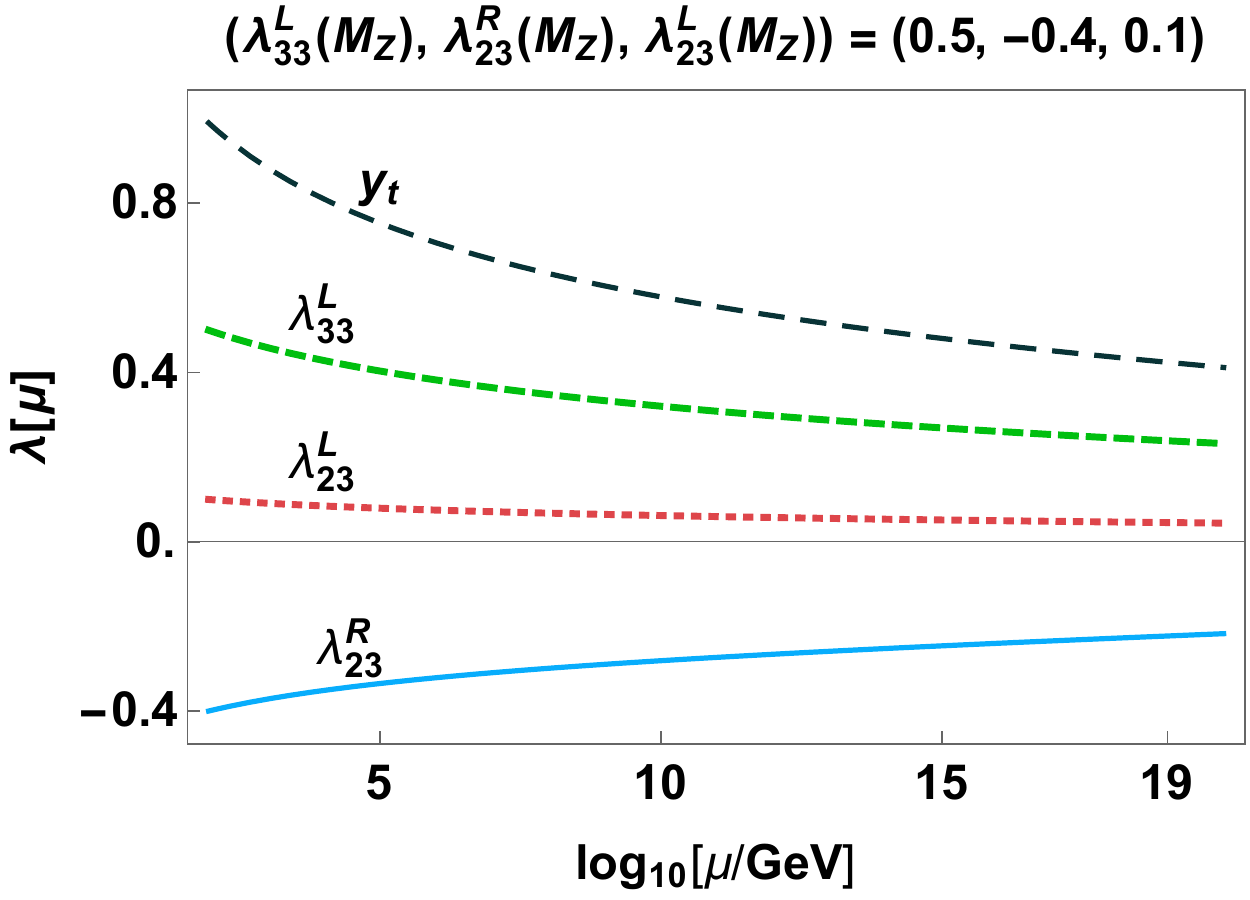}\label{fig:Yukawaa}}&
\subfloat[\quad\quad\quad(b)]{\includegraphics[height=4.5cm,width=5.4cm]{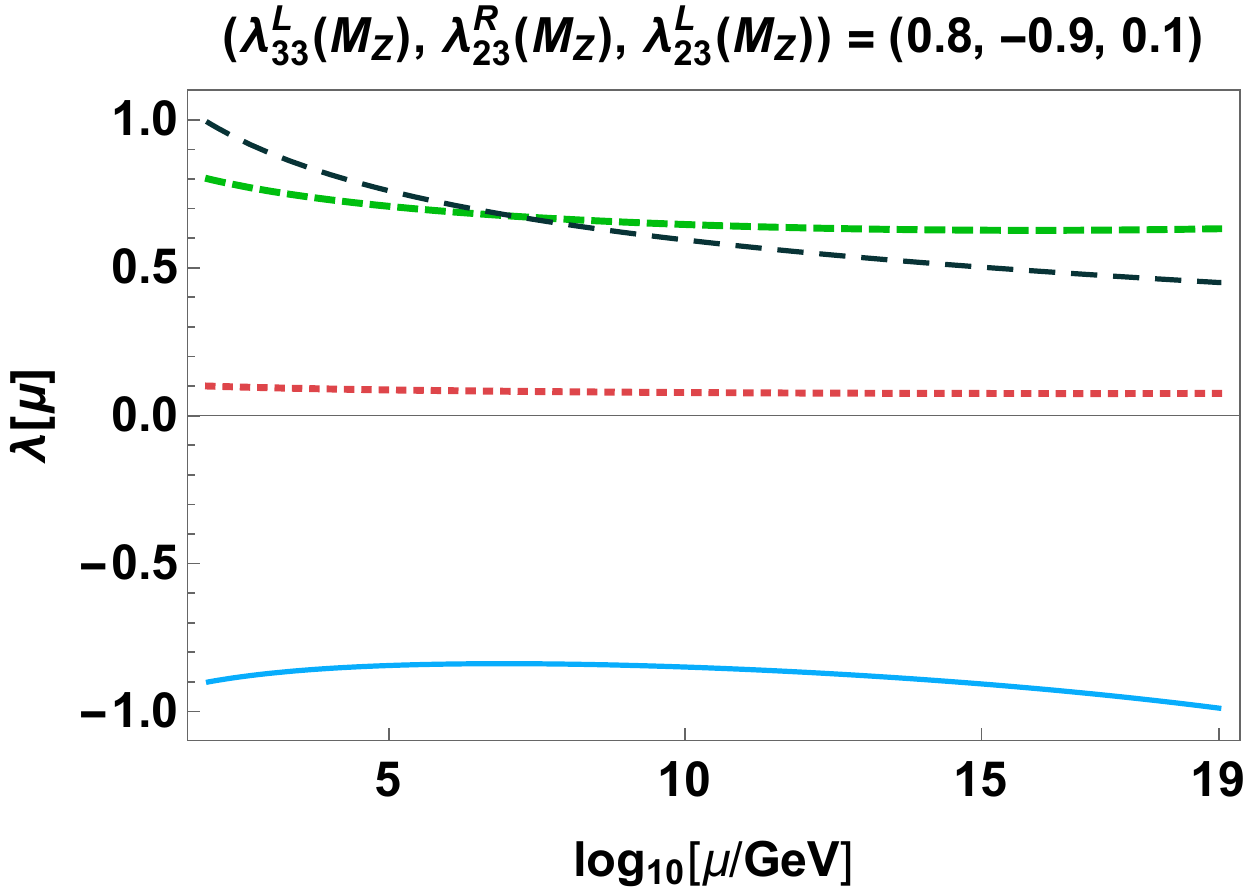}\label{fig:Yukawab}}&
\subfloat[\quad\quad\quad(c)]{\includegraphics[height=4.5cm,width=5.4cm]{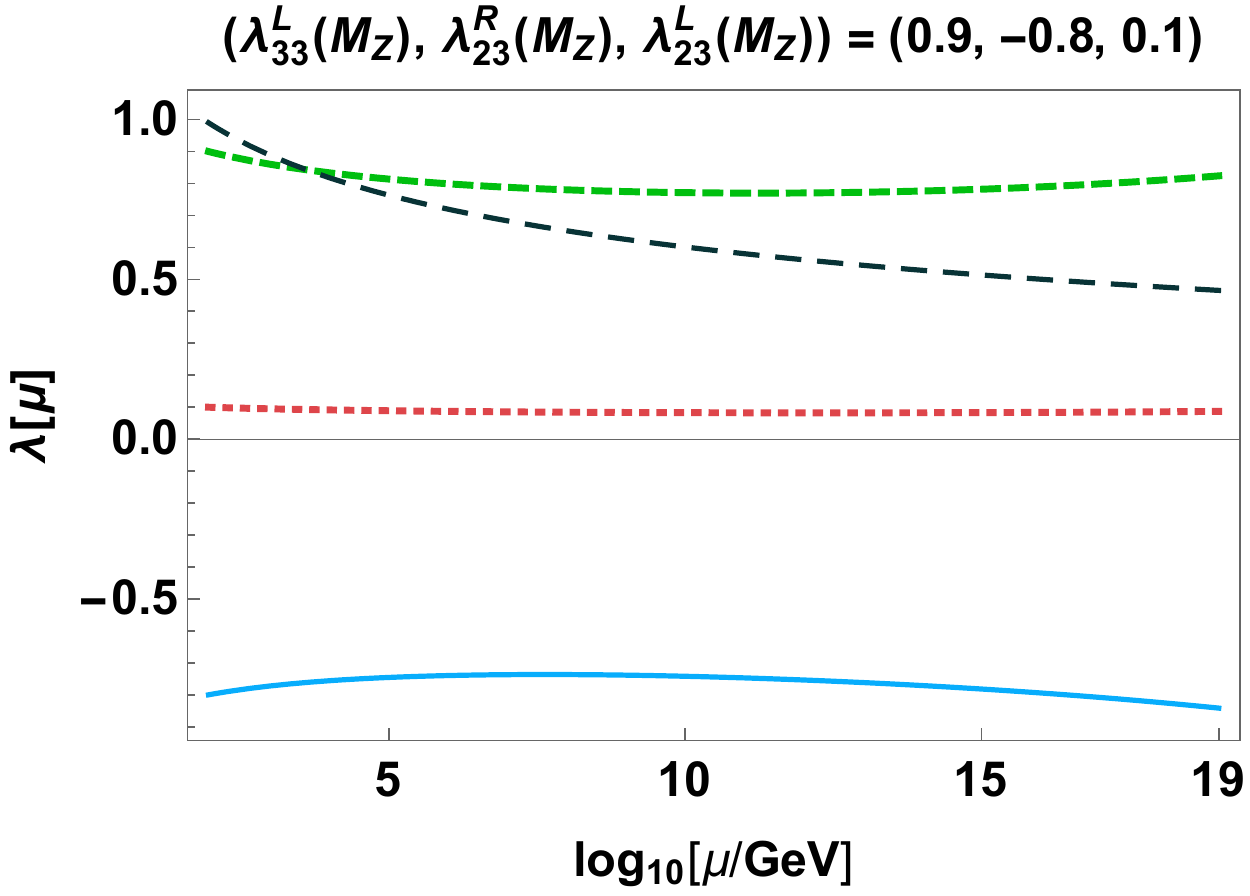}\label{fig:Yukawac}}\\\\
\subfloat[\quad\quad\quad(d)]{\includegraphics[height=4.5cm,width=5.4cm]{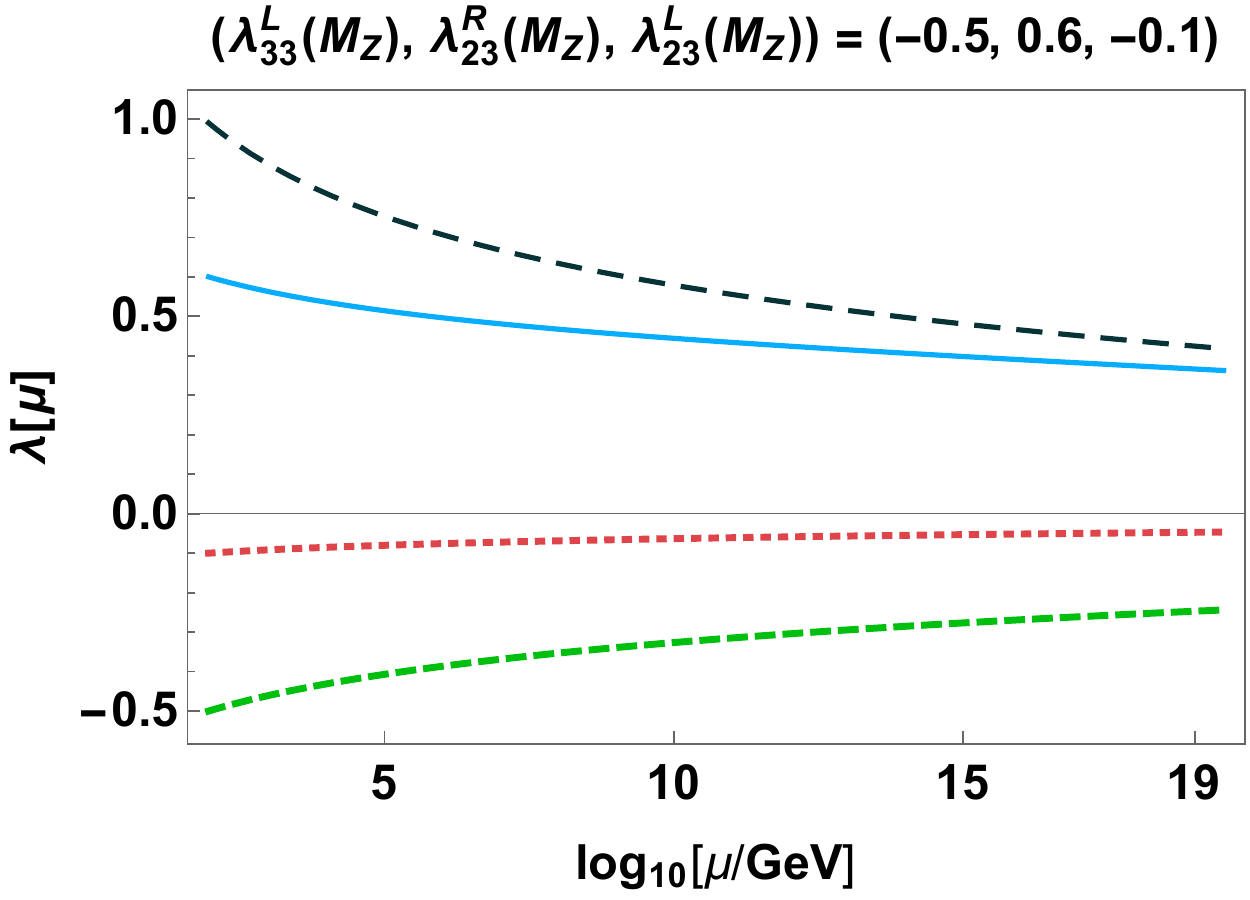}\label{fig:Yukawad}}&
\subfloat[\quad\quad\quad(e)]{\includegraphics[height=4.5cm,width=5.4cm]{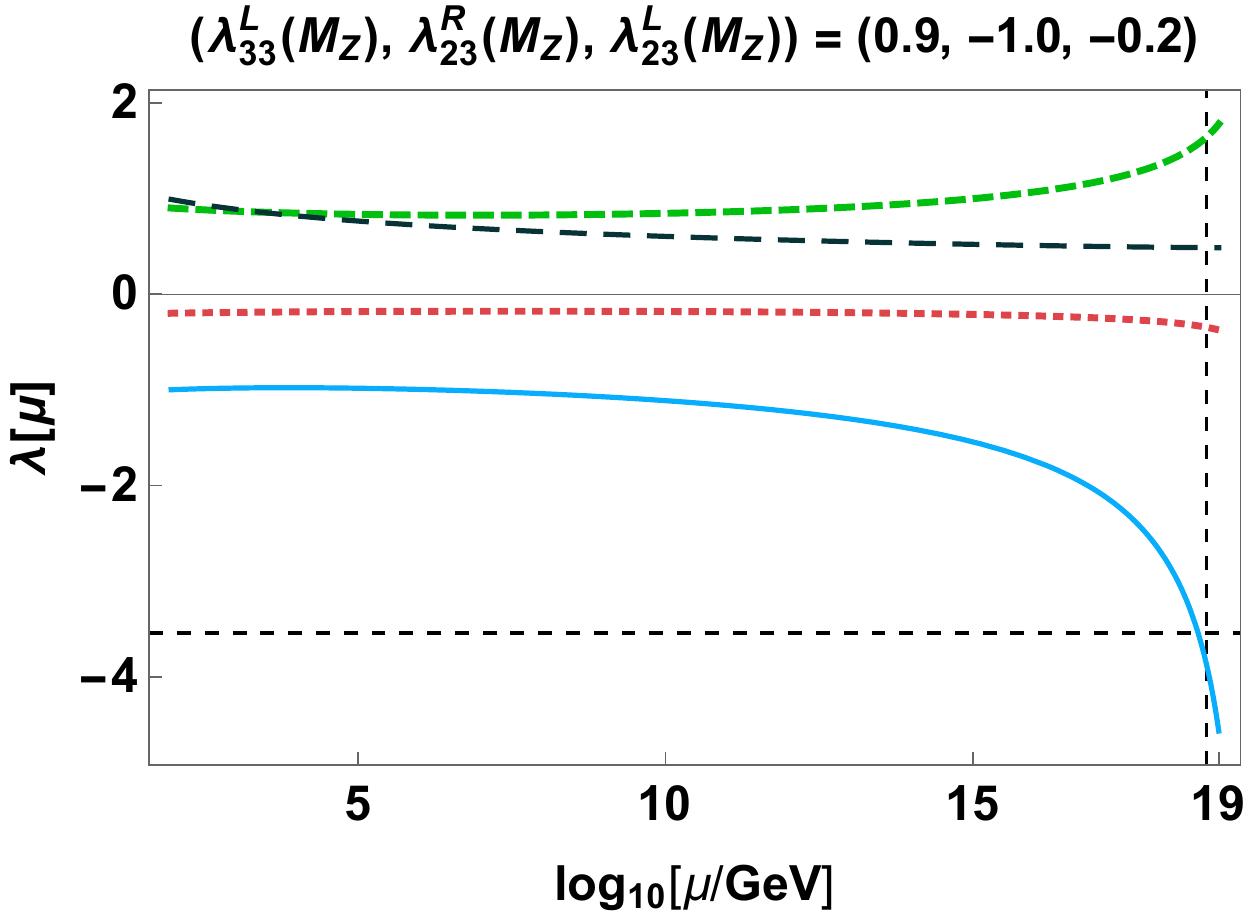}\label{fig:Yukawae}}&
\subfloat[\quad\quad\quad(f)]{\includegraphics[height=4.5cm,width=5.4cm]{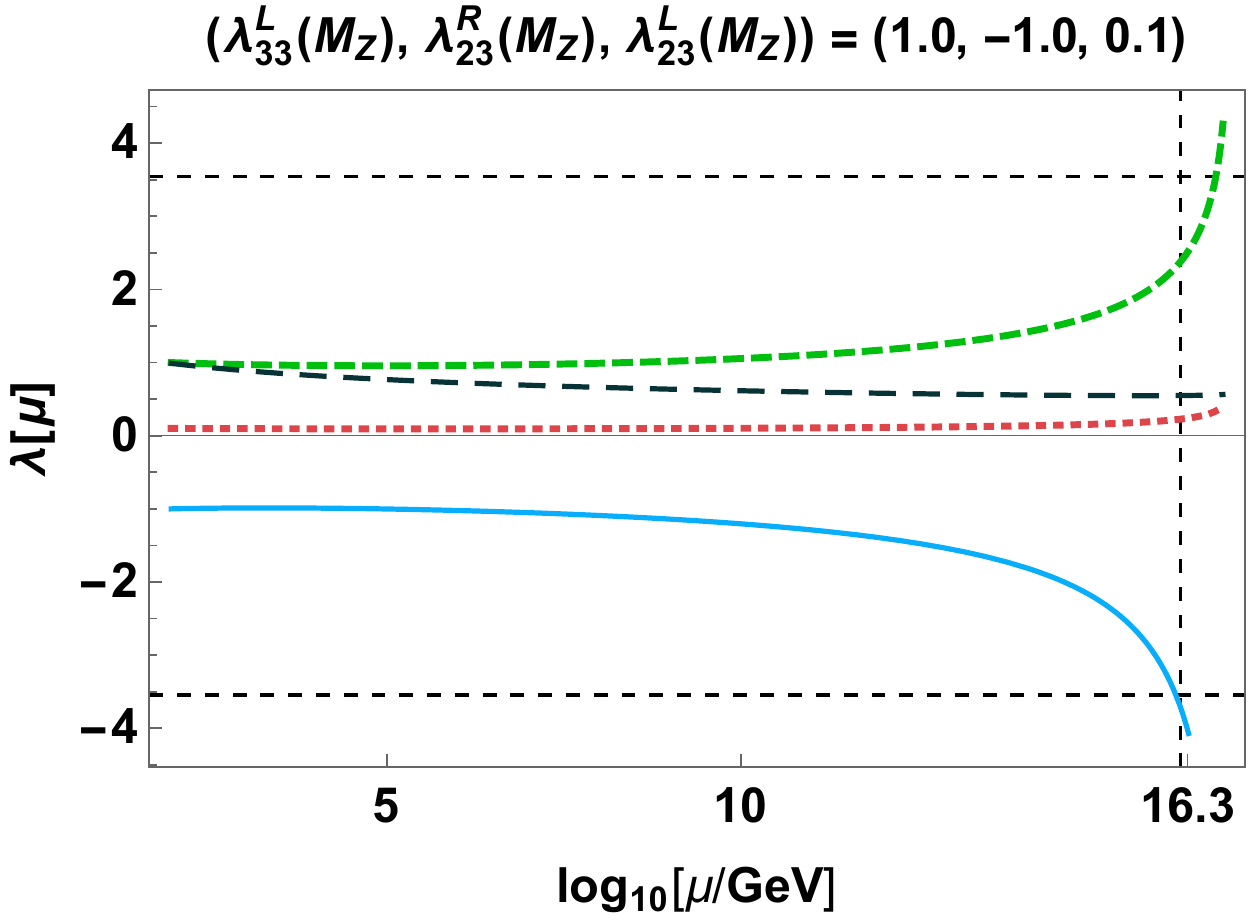}\label{fig:Yukawaf}}\\\\
\subfloat[\quad\quad\quad(g)]{\includegraphics[height=4.5cm,width=5.4cm]{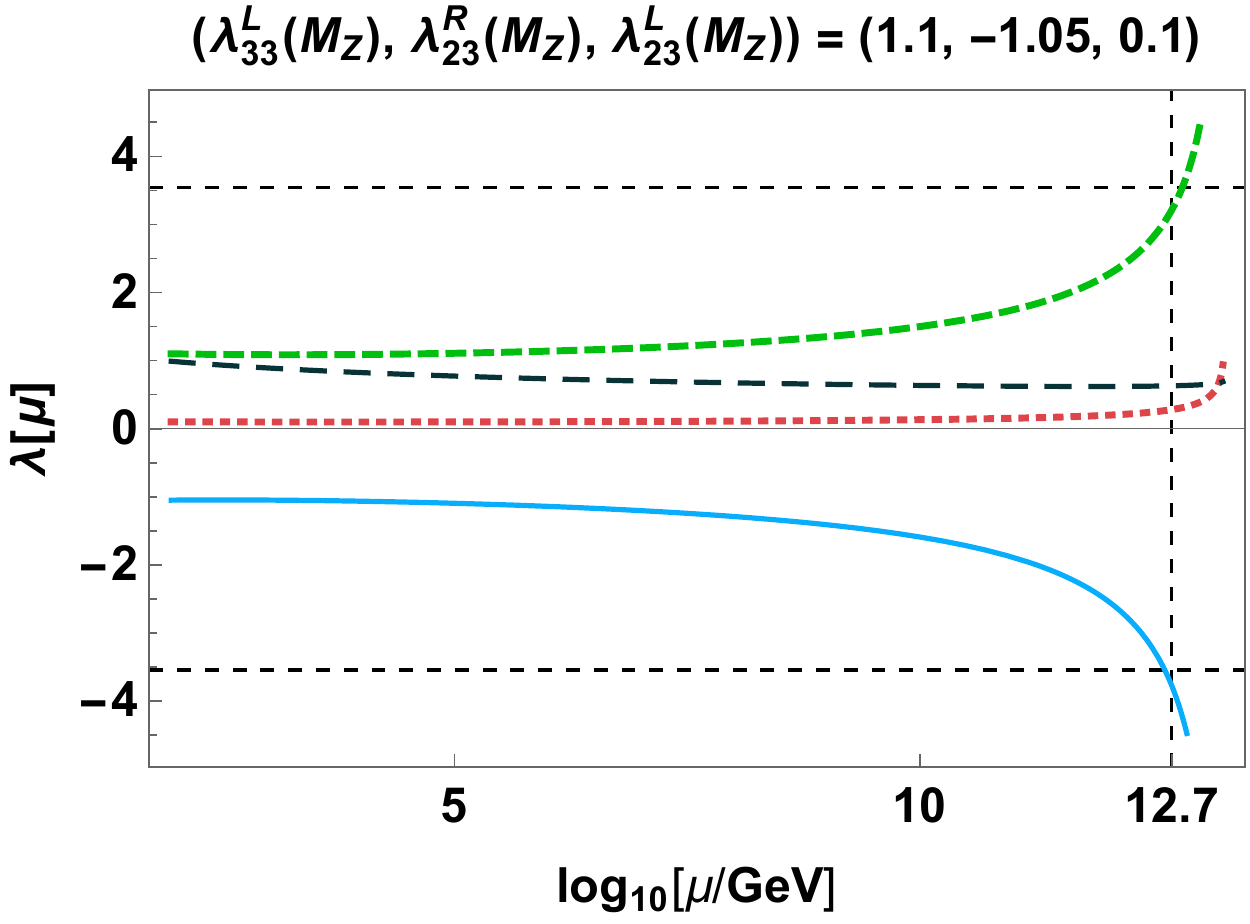}\label{fig:Yukawag}}&
\subfloat[\quad\quad\quad(h)]{\includegraphics[height=4.5cm,width=5.4cm]{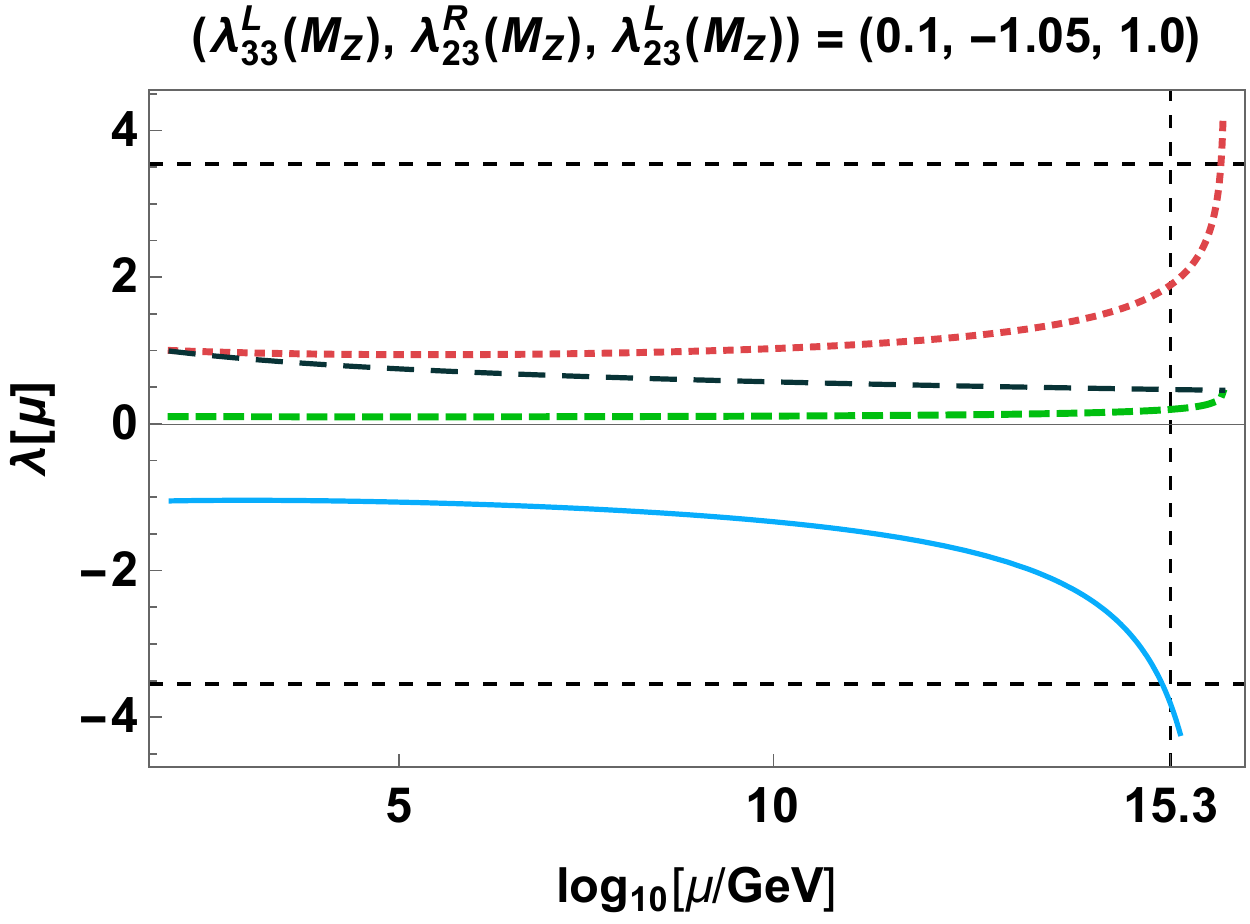}\label{fig:Yukawah}}&
\subfloat[\quad\quad\quad(i)]{\includegraphics[height=4.5cm,width=5.4cm]{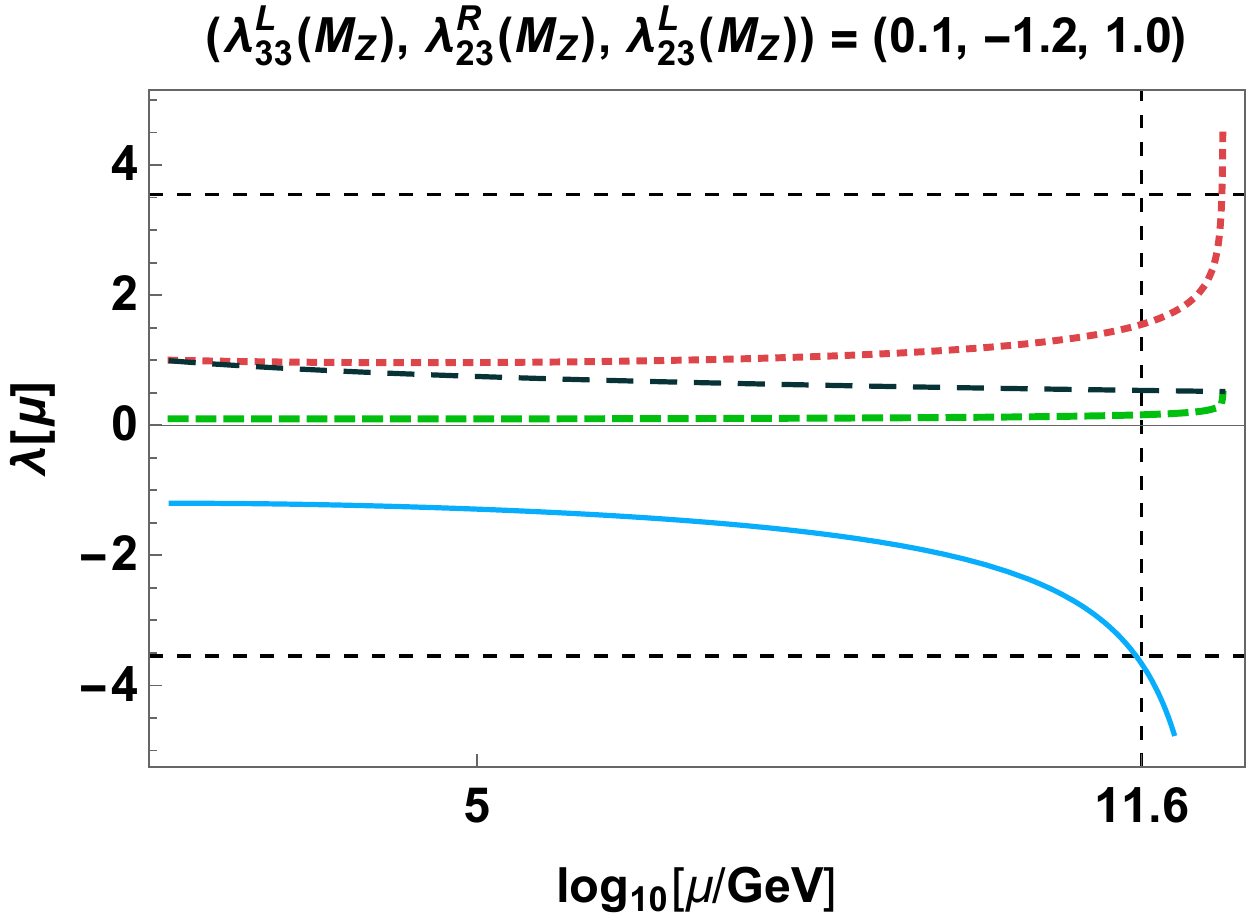}\label{fig:Yukawai}}\\\\
\end{tabular}
\vspace{-0.5cm}\caption{Behaviour of Yukawa couplings with various benchmark values at the EW scale. The labels of the couplings are given in the first plot. The dashed horizontal lines denote the values of the assumed perturbativity bound, $\pm\sqrt{4\pi}$. The dashed vertical line denotes the energy scale at which this bound is first reached.}
\label{fig:Yukawa}
\end{figure*}

\subsection{RG running of the Yukawa couplings and perturbativity}\label{sec:Yukawa}
\noindent
One of the questions raised by the introduction of new Yukawa couplings is whether the new model remains perturbative up to high-enough energies. This is particularly important in the GUT framework since the RG running of the gauge couplings is performed under the assumption of perturbativity. Fortunately, with the latest LHC data, we have quite a large available parameter space in the deep perturbative region, as can be seen in Fig.~\ref{fig:parascan}. While this suggests that the model is safe in terms of perturbativity as long as the new Yukawa couplings are small enough at the electroweak scale, it is still informative to investigate the RG running in detail,  especially the case in which the Yukawa couplings take larger initial values. We address this issue in this part of the paper\footnote{A perturbativity analysis for the case of the Standard Model augmented by a leptoquark $S_1$ was done in Ref.~\cite{Bandyopadhyay:2016oif}, as well. Note that their Yukawa matrix is flavour diagonal, and hence different from the one in this paper.}.

The equations for the RG running of the Yukawa couplings are given in Appendix~\ref{sec:Appendix}. Results for various benchmark cases are displayed in Fig.~\ref{fig:Yukawa}. Among the $S_1$ mass values we have considered in this paper, the most constrained parameter space is that of $M_{S_1}=2$ TeV. Therefore, we choose our benchmark values from the parameter space for this mass value, displayed in Fig.~\ref{fig:parascan}. Note that $\lambda^{L}_{33}(M_Z)$ and $\lambda^{L}_{23}(M_Z)$ cannot both be large and that $\lambda^{L}_{33}(M_Z)$ and $\lambda^{R}_{23}(M_Z)$ must have opposite signs, as can be seen in Fig.~\ref{fig:parascan}. When $\lambda^{L}_{23}(M_Z)$ is taken in the interval $[-0.2, 0.1]$, the system remains perturbative up to the grand unification scale, $\sim 10^{15}$-$10^{17}$ GeV, even when $|\lambda^{L}_{33}(M_Z)|$, $|\lambda^{R}_{23}(M_Z)|\approx 1$, as displayed in the first six plots in Fig.~\ref{fig:Yukawa}. However, it deteriorates quickly for larger values of $|\lambda^{L}_{33}(M_Z)|$ and $|\lambda^{R}_{23}(M_Z)|$ [Fig.~\ref{fig:Yukawag}]. When $|\lambda^{L}_{23}(M_Z)|$ is large, the parameter space is quite limited for  $|\lambda^{L}_{33}(M_Z)|$, which is in the $[0.10, 0.15]$ band (Fig.~\ref{fig:parascan}). In this case as well, the system is well behaved up to $|\lambda^{L}_{23}(M_Z)|$, $|\lambda^{R}_{23}(M_Z)|\approx 1$ [e.g. Fig.~\ref{fig:Yukawah}], and the situation declines for the values above in that the perturbativity bound is reached below the unification scale [e.g. Fig.~\ref{fig:Yukawai}].

Note that although we perform the Yukawa RG running based on the SM augmented with a TeV scale leptoquark $S_1$ all the way up the UV, these equations are prone to changes above the intermediate symmetry-breaking scale provided that there is one (as in the examples studied in Sec.~\ref{sec:unification}), mainly due to contributions from the running of the scalars whose masses are around the intermediate scale. However, these effects are expected to be minor due to the corresponding beta-function coefficients not being large enough to significantly change the logarithmic RG running~\cite{Altarelli:2013aqa}. This is even more likely to be the case especially if this symmetry-breaking scale, for instance the scale in our scenario where the Pati-Salam symmetry is spontaneously broken to the symmetry of the SM, is considerably close to the scale of the $\mathrm{SO}(10)$ symmetry-breaking scale (as in the second case in Sec.~\ref{sec:unification}, namely model $A_2$), suggesting that the supposed modification in the Yukawa running is indeed not an issue of concern since the slow logarithmic running would most likely does not significantly alter the outcome in this small interval. Threshold corrections due to the intermediate scale are also known to be subleading to the one-loop running~\cite{Babu:2016bmy}.

Furthermore, although we have studied some specific examples for gauge coupling unification, there are no restrictions on the choice of the mass values of the high-energy field content regarding which fields remain heavy at the unification scale and which ones slide through the intermediate scale, as long as the gauge coupling unification is realized (and as long as the intermediate scale is high enough to evade the proton-decay constraints for the terms not forbidden by any symmetry in the Lagrangian). The main point of our case that we emphasize is that if an $\mathrm{SO}(10)$ theory is indeed the UV completion of the SM, then one may naturally anticipate a TeV scale leptoquark accompanying the SM Higgs field, and this could define the field content up to very high energies. Beyond that, one has the freedom to choose the high-energy particle content and the corresponding potential terms in the Lagrangian that lead to an appropriate symmetry-breaking sequence in which the Yukawa couplings remain in the perturbative realm above the intermediate symmetry-breaking scale. With this in mind, even the other cases, displayed in Fig.~\ref{fig:Yukawag} and Fig.~\ref{fig:Yukawai}, that suffer from perturbativity problem at relatively low energies could arguably get a pass, as long as the high energy content of the theory is chosen such that the intermediate symmetry-breaking occurs before the perturbativity bound is reached and such that the couplings remain in the perturbative realm up to the unification scale.
 
 We comment in passing on the unification-scale implications of our model regarding the fermion mass spectrum. It has been known in the literature that obtaining a realistic Yukawa sector in the $\mathrm{SO}(10)$ framework is not trivial. None of the single- and dual-field combinations of the scalar fields $\mathbf{10}_H$, $\mathbf{120}_H$, and $\mathbf{126}_H$ yields GUT-scale relations between fermion masses consistent with the SM values~\cite{Bajc:2005zf,Babu:2016bmy}. On the other hand, it was concluded in Ref.~\cite{Babu:2016bmy} that a Higgs sector consisting of a real $\mathbf{10}_H$, a real $\mathbf{120}_H$, and a complex $\mathbf{126}_H$ can provide a fermion mass spectrum at the unification scale that matches the expected values obtained by the RG running of the SM with small threshold corrections at the intermediate scale. Since this is the scalar sector we adopt in this paper, it is informative to inspect whether the light leptoquark $S_1$ can register significant changes to the expected fermion mass values at the unification scale obtained via the SM RG running.
 
The results, obtained by using the equations given in Appendix~\ref{sec:Appendix} and the same values for the input parameters at $M_Z$ as in Ref.~\cite{Babu:2016bmy}, are given in Table~\ref{tab:fermionmass}, some of which are displayed in terms of mass ratios relevant in the $\mathrm{SO}(10)$ framework. The unification scale is selected as $M_U=2\times10^{16}$ GeV, which is around the exponential midpoint of the unification scales of our models $A_1$ and $A_2$ discussed in Sec.~\ref{sec:unification}; small numerical differences depending on the $M_U$ value in that range is not relevant to our discussion here. Our values for the fermion masses at $M_U$ in the SM case differ from the ones in Ref.~\cite{Babu:2016bmy} by $1$-$5~\%$, which might be due to a combination of effects coming from the running of the right-handed neutrinos above the intermediate scale and the threshold corrections, both of which are ignored in our estimation.\footnote{We do not discuss neutrino masses in this paper but have no reason to suspect that the outcome would be different than Ref.~\cite{Babu:2016bmy} particularly because their intermediate symmetry-breaking scales at which the generation of neutrino masses through the seesaw mechanism occurs are quite close to the ones in our models $A_1$ and $A_2$.} The results for the leptoquark case are given for three benchmark points in the $(\lambda_{33}^L,\lambda_{23}^R,\lambda_{23}^L)$ parameter space that are consistent with the perturbativity analysis above. As expected, when all three  of the  new Yukawa couplings are in the deep perturbative region, the differences from the SM values remain insignificant. When the couplings get larger close to unity, some deviations are observed yet they do not become substantial.  Therefore, we conclude that the addition of $S_1$ to the particle content up to the TeV scale does not lead to significant changes regarding the expected fermion masses at the unification scale and therefore the analysis of Ref.~\cite{Babu:2016bmy}, based on using the extrapolated SM values at the unification scale as input data in order to numerically fit them to the parameters of the Yukawa sector of their $\mathrm{SO}(10)$ model, can be applied in our case as well.
 
\begin{table*}[t!]
\caption{Fermion masses at the unification scale $M_U=2\times 10^{16}$ GeV that are selected for the sake of argument as the exponential midpoint of the unification scales of the high-energy scenarios discussed in Sec.~\ref{sec:unification}. Clearly, inclusion of $S_1$ in the low-energy particle spectrum does not lead to significant changes in the fermion mass values at the unification scale compared to the SM predictions, especially in the deep perturbative region in the $(\lambda_{33}^L,\lambda_{23}^R,\lambda_{23}^L)$ parameter space.}
\begin{center}
\begin{tabular}{c||c|c|c|c}
\hline
$\vphantom{\Big|}$
\diaghead{\theadfont Diag ColumnmnHead II}%
  {Fermion masses/ratios}{$(\lambda_{33}^L,\lambda_{23}^R,\lambda_{23}^L)$} & $\; (0,0,0);\;\textrm{SM}\;\;\;$  &$(0.5, -0.4, 0.1)$ & $(0.8, -0.9, 0.1)$ & $(0.1, -1.0, 1.0)$ \\
\hline\hline
$\vphantom{\big|}$
$m_t/m_b$  & $75.57$ & $75.66 $  & $76.37$ & $75.36$\\
$\vphantom{\big|}$
$m_{\tau}/m_b$ 	& $1.63$		& $1.74$ & $2.31$ & $4.28$\\
$\vphantom{\big|}$
$m_{\mu}/m_s$ 		& $4.41$	& $4.46$   & $4.46$ & $3.94$\\
$\vphantom{\big|}$
$m_{e}/m_d$ 	& $0.400$& $0.401$ & $0.401$ & $0.401$ \\
$\vphantom{\big|}$
$m_t/{\footnotesize\textrm{GeV}}$  &      $79.32$&                $79.52$     &   $83.46$      & $78.14$\\
$\vphantom{\big|}$
$m_c/{\footnotesize\textrm{GeV}}$ 	 & $0.254$&           $0.254$       &     $0.275$     &  $0.344$\\
$\vphantom{\big|}$
$m_{\mu}/(10^{-3} {\footnotesize\textrm{GeV}})$ 	& $100.505$       & $100.576$    & $101.518$      & $100.175$\\
$\vphantom{\big|}$
$m_{e}/(10^{-3} {\footnotesize\textrm{GeV}})$ 	  & $0.476$             & $0.476$    &   $0.481$      & $0.475$ \\
$\vphantom{\big|}$
$m_u/(10^{-3} {\footnotesize\textrm{GeV}})$ 	   & $0.469$             & $0.464$     & $0.469$       & $0.462$ \\
\hline
\end{tabular}
\label{tab:fermionmass}
\end{center}
\end{table*}


\section{Summary and Discussions}
\label{sec:outlook}
\noindent
In this paper, we considered the scenario that there is a single scalar leptoquark, $S_1$, at the TeV scale, and it is the colour-triplet component of a real $\mathbf{10}$ of $\mathrm{SO}(10)$, which also contains a $\mathrm{SU}(2)$ doublet which is identified as the SM Higgs. In this scenario, the leptoquark being the only scalar entity other than the SM Higgs is natural; a peculiar mass splitting between the components of $\mathbf{10}$ does not occur and the leptoquark picks up an electroweak-scale mass together with the SM Higgs, as expected. This is appealing because this leptoquark by itself can potentially explain the $B$-decay anomalies at the LHC~\cite{Bauer:2015knc,Becirevic:2016oho,Cai:2017wry,Freytsis:2015qca}. 

The $\mathrm{SO}(10)$ grand unification framework is an appealing scenario, which has been heavily studied in the literature~\cite{Fritzsch:1974nn,Chang:1983fu,Chang:1984uy,Chang:1984qr,Deshpande:1992au,Bajc:2005zf,Bertolini:2009qj,Babu:2012vc,Altarelli:2013aqa,Aydemir:2015oob,Aydemir:2016qqj,Babu:2015bna,Babu:2016bmy}. It unifies the three forces in the SM, explains the quantization of electric charge, provides numerous dark matter candidates, accommodates seesaw mechanism for small neutrino masses, and justifies the remarkable cancellation of anomalies through the anomaly-free nature of the $\mathrm{SO}(10)$ gauge group. Moreover, the fermionic content of the SM fits elegantly in $\mathbf{16}_F$, including a right-handed neutrino for each family. The particles in the SM (except probably the neutrinos) acquire masses up to the electroweak scale through interacting with the electroweak-scale Higgs field, which is generally assigned to a real $\mathbf{10}_H$. Considering that the leptoquark $S_1$ is the only other component in $\mathbf{10}_H$,  a TeV scale $S_1$, from this perspective, is consistent with the idea that it is the last piece of the puzzle regarding the particle content up to the electroweak scale (modulo the right-handed neutrinos). Therefore, the possible detection of $S_1$, in the absence of any other new particles, at the TeV scale could be interpreted as evidence in favour of $\mathrm{SO}(10)$ grand unification.  

One obvious issue of concern is proton decay since the leptoquark $S_1$ possesses the right quantum numbers for it to couple to potentially dangerous diquark operators.  On the other hand, the proton stability could possibly be ensured through various mechanisms~\cite{Cox:2016epl,Bajc:2005zf,Pati:1974yy,Bauer:2015knc,Dvali:1995hp,Aydemir:2018cbb}. In this paper, we ensured the proton stability by assuming a discrete symmetry that is imposed in an \emph{ad hoc} manner below the Pati-Salam breaking scale. It would certainly be more compelling to realize a similar mechanism at the fundamental level although it seems unlikely that it would interfere with the bottom line of this work.   

Having a single $S_1$ leptoquark at the TeV scale as the only new physics remnant from our $\mathrm{SO}(10)$ GUT model,
we investigated how competent this $S_1$ is at addressing the $R_{D^{(*)}}$ anomalies while simultaneously satisfying other
relevant constraints from flavour, electroweak and the direct LHC searches. We adopted a specific Yukawa coupling texture
with only three free (real) nonzero couplings viz. $\lm_{23}^L$, $\lm_{33}^L$ and $\lm_{23}^R$. We have found that this
minimal consideration can alleviate the potential tension between the $R_{D^{(*)}}$-favoured region and the $R_{K^{(*)}}^{\nu\nu}$
measurements. The $Z\to\tau\tau$ decay constrains $\lm_{33}^L$ whereas the $\tau\tau$ resonance
search at the LHC puts complimentary bounds on $\lm_{23}^L$ and $\lm_{23}^R$. By combining these constraints with all the relevant flavour constraints coming from the latest data on $R_{D^{(*)}}$, $F_L{(D^*)}$, $P_\tau(D^*)$, $R_{K^{(*)}}^{\n\n}$ we
have found that a substantial region of the $R_{D^{(*)}}$-favoured parameter space is allowed. Our multiparameter analysis clearly shows that contrary to the common perception, a single leptoquark solution to the observed $R_{D^{(*)}}$ anomalies with $S_1$ is still a viable solution.

Evidently, by introducing new degrees of freedom into our framework, one can, \emph{a priori}, enlarge the allowed
parameter region. For example, by considering some of the couplings as complex or by choosing a complex (instead of a real) 
$\mathbf{10}$ representation of 
$\mathrm{SO}(10)$, which will introduce an additional $S_1$ and another complex scalar doublet at the TeV scale, one can
relax our obtained bounds. We also pointed out search strategies for $S_1$ at the LHC using symmetric and asymmetric pair- and single-production channels.
 Systematic studies of these channels are discussed elsewhere~\cite{Chandak:2019iwj}.


\section*{Acknowledgements}
\noindent
We thank Diganta Das for valuable discussions and collaboration in the early stages of the project.~Work of U.A. is supported in part by the Chinese Academy of Sciences President's International Fellowship Initiative (PIFI) under Grant No.~2020PM0019; the Institute of High Energy Physics, Chinese Academy of Sciences, under Contract No.~Y9291120K2; and the National Natural Science Foundation of China (NSFC) under Grant No.~11505067. T.M. is financially supported by the Royal Society of Arts and Sciences of Uppsala
as a researcher at Uppsala University and by the INSPIRE Faculty Fellowship of the
Department of Science and Technology (DST) under grant
number IFA16-PH182 at the University of Delhi. S.M. acknowledges support from the Science and Engineering Research Board (SERB), DST, India under grant number ECR/2017/000517.
\appendix
\section{Renormalization group running of the Yukawa couplings}\label{sec:Appendix} \noindent
The new Yukawa matrices in the Lagrangian given in Eq.~\eqref{eq:lagrangianLQ} are taken in our setup as\\
\begin{eqnarray}
\Lambda^{L} \rightarrow \left(\begin{array}{ccc}0 & 0 & 0 \\0 & 0 & \lambda^{L}_{23} \\0 & 0 & \lambda^{L}_{33} \end{array}\right)~ \mbox{and}~ \Lambda^{R} \rightarrow \left(\begin{array}{ccc}0 & 0 & 0 \\0 & 0 & \lambda^{R}_{23} \\0 & 0 &0 \end{array}\right).\qquad\nonumber\\
\end{eqnarray}
Following Ref.~\cite{Machacek:1983fi} (or implementing the model in SARAH~\cite{Staub:2012pb}), the corresponding one-loop RG equations can be found as
\begin{eqnarray}
16 \pi^2 \beta_{\lambda^R_{23}}&=& 3\left(\lambda^{R}_{23}\right)^3-\lambda^{R}_{23}\left(\frac{13}{3}g_1^2+4g_3^2\right)\nn\\
&&+2 \lambda^{R}_{23}\left( \left(\lambda^L_{23}\right)^2+\left(\lambda^L_{33}\right)^2\right),\\
16 \pi^2 \beta_{\lambda^L_{33}}&=& 4 \left(\lambda^L_{33}\right)^3-\lambda^L_{33}\Big(-\frac{y_t^2}{2}+\frac{5}{6}g_1^2+\frac{9}{2}g_2^2\nn\\
&&+4g_3^2\Big)+\lambda^L_{33}\left( \left(\lambda^R_{23}\right)^2+4\left(\lambda^L_{23}\right)^2\right),\end{eqnarray}
\begin{eqnarray}
16 \pi^2 \beta_{\lambda^L_{23}}&=& 4 \left(\lambda^L_{23}\right)^3-\lambda^L_{23}\left(\frac{5}{6}g_1^2+\frac{9}{2}g_2^2+4g_3^2\right)\nn\\
&&+\lambda^L_{23}\left( \left(\lambda^R_{23}\right)^2+4\left(\lambda^L_{33}\right)^2\right),\\
16 \pi^2 \beta_{y_t}&=&3 y_t^3+y_t\left(\frac{3}{2} y_t^3-\frac{17}{12} g_1^2-\frac{9}{4} g_2^2-8 g_3^2\right)\nn\\
&&+\frac{1}{2} y_t \left(\lambda^L_{33}\right)^2,\\
16 \pi^2 \beta_{y_c}&=&y_c\left(3y_t^2-\frac{17}{12} g_1^2-\frac{9}{4} g_2^2-8 g_3^2\right)\nn\\
&&+\frac{1}{2} y_c \left(\left(\lambda^R_{23}\right)^2+\left(\lambda^L_{23}\right)^2\right),\\
16 \pi^2 \beta_{y_u}&=&y_u\left(3y_t^2-\frac{17}{12} g_1^2-\frac{9}{4} g_2^2-8 g_3^2\right),\\
16 \pi^2 \beta_{y_b}&=&y_b\left(\frac{3}{2}y_t^2-\frac{5}{12} g_1^2-\frac{9}{4} g_2^2-8 g_3^2\right)\nn\\
&&+\frac{1}{2} y_b \left(\lambda^L_{33}\right)^2,
\end{eqnarray}
\begin{eqnarray}
16 \pi^2 \beta_{y_s}&=&y_s\left(3y_t^2-\frac{5}{12} g_1^2-\frac{9}{4} g_2^2-8 g_3^2\right)\nn\\
&&+\frac{1}{2} y_s \left(\lambda^L_{23}\right)^2,\\
16 \pi^2 \beta_{y_d}&=&y_d\left(3y_t^2-\frac{5}{12} g_1^2-\frac{9}{4} g_2^2-8 g_3^2\right),\\
16 \pi^2 \beta_{y_e}&=&y_e\left(3y_t^2-\frac{15}{4} g_1^2-\frac{9}{4} g_2^2\right),\\
16 \pi^2 \beta_{y_{\mu}}&=&y_{\mu}\left(3y_t^2-\frac{15}{4} g_1^2-\frac{9}{4} g_2^2\right),
\end{eqnarray}
\begin{eqnarray}
16 \pi^2 \beta_{y_{\tau}}&=&y_{\tau}\left(3y_t^2-\frac{15}{4} g_1^2-\frac{9}{4} g_2^2\right)\nn\\
&&+\frac{3}{2} y_{\tau} \left(\left(\lambda^R_{23}\right)^2+\left(\lambda^L_{23}\right)^2+\left(\lambda^L_{33}\right)^2\right),\qquad
\end{eqnarray}
%
where $\beta_{y}\equiv \mu\dfrac{d y}{d\mu}$ and the RG running of the gauge couplings are performed in the usual way according to Eq.~(\ref{gaugerunning}). In the Yukawa running above, only the dominant terms are taken into account since the subleading ones are significantly suppressed and have no noticeable effects on our analysis.

\newpage
\bibliography{reference}{}
\bibliographystyle{JHEPCust}

\end{document}